\documentstyle[12pt]{article}
\begin{document}
\title{Initial bound state studies in light-front QCD.}
\author{ Martina Brisudov\' a  and Robert Perry \\
{\it Department of Physics}\\ {\it The Ohio State University,
 Columbus, OH 43210.} }
\date{}
\maketitle
\abstract{
We present the first numerical QCD bound state calculation based on a
renormalization
group-improved light-front Hamiltonian formalism. The QCD Hamiltonian is
determined to second order in the coupling, and it includes two-body confining
interactions. We make a momentum expansion, obtaining an equal-time-like
Schr{\H{o}}dinger equation. This is solved for quark-antiquark constituent
states, and we obtain a set of self-consistent parameters by fitting B meson
spectra.
}
\vskip .25in
\newpage
\section{Introduction.}
Recently,  a new approach to renormalization and solving for QCD
bound states has been advocated
%by Wilson and others
\cite{thelongpaper,P2,POLAND1, POLAND2}.
The goal
 is to build a bridge between QCD and a constituent quark model (CQM).
It has been
argued that it is convenient to use a light-front formulation of the theory,
because on the light-front it is possible to make the vacuum trivial simply
by  implementing a small longitudinal cutoff. As a result
 all partons in a hadronic state
are connected to the hadron, instead of being disconnected excitations in a
complicated medium. The price to pay is a considerably more complicated
renormalization problem.

In this paper we briefly describe this new approach, concentrating on
aspects, necessary to appreciate the simple calculation we present. The
calculations of the heavy meson spectra, for systems containing one heavy
and one light quark, is intended to elucidate the approach and provide
qualitative tests of the leading terms revealed in the effective
Hamiltonian by the renormalization group. Only the first step in this
calculation is taken here, and later work will focus on the spin-dependent
structure. We refer the reader to literature for many details
\cite{thelongpaper,P2,POLAND1,POLAND2}.

The new approach consists
of two steps. The first step is renormalization. The second step is a bound
state calculation.

 The aim of the first step is to find an effective renormalized
 Hamiltonian at hadronic energy
scales starting with a Hamiltonian which is consistent with perturbative
 QCD at high energy scales. A natural starting point is the
 canonical light-front
QCD Hamiltonian, although it cannot be complete --- there must  be other
operators, which cannot be determined from the perturbative
 behaviour of the theory.
G{\l}azek and Wilson \cite{similarity}
 designed a similarity transformation to lower the
cutoff scale, which is tailored to
make the Hamiltonian look more like a CQM Hamiltonian (see, for example,
\cite{CQM} and references in \cite{thelongpaper}).
The cutoff violates manifest gauge invariance and Lorentz covariance,
 and thus these symmetries are no longer a guide to
what operators are allowed in the Hamiltonian.

Similarity transformations can be  designed to bring the Hamiltonian toward a
band-diagonal form by eliminating matrix elements between states which differ
drastically in light-front energy. Effects of couplings that are removed have
to be put
directly into the
Hamiltonian as new effective interactions. One important consequence is that
two-body
potentials are generated. In fact, the similarity transformation generates a
logarithmic confining potential already at order $g^2$ \cite{P2,robpoland}.
 This part
of the calculation is done perturbatively. If the similarity
transformation can be done analytically, it is easy to use a powerful method
called coupling coherence \cite{pertren,couplcoh}
to determine all new  terms.
We will illustrate this
procedure, which is straightforward to second order, in the following sections.

The second step is the bound state calculation itself. The effective
Hamiltonian is divided
into $H_0$, a part which is solved
nonperturbatively, and $V$, the difference between the original Hamiltonian
and  $H_0$. The effects of $V$ are to be computed using bound state
perturbation theory. The criteria for choosing $H_0$ are that  it
approximates the physics relevant for hadronic bound states
 as closely as possible (we take a hint from the CQM and
include constituent masses and two-body potentials) and yet it must be
 manageable.
We emphasize that the approach is tailored within limits
to take advantage of the
successful phenomenology, but it does not stop there.
We can systematically improve the calculations, both by computing
corrections to the Hamiltonian and higher higher terms in bound state
perturbation theory.
For example, any terms added by hand (e.g. constituent masses) can be added
 in such a way that at the physical value of
the coupling they %approximate interactions
reduce to terms in the effective QCD Hamiltonian
(for details see \cite{thelongpaper,POLAND2}).

In this paper we present one of the simplest possible calculation of QCD
bound states based on the new approach. In the first step we find the effective
Hamiltonian to order $g^2$ using a similarity transformation and coupling
coherence, and in the second step we solve for the lowest lying
$q\bar{q}$ color singlet states with arbitrary but nonzero masses in the
nonrelativistic limit. These approximations are severe, but we will see
that the qualitative results are good.
We will not explicitly show operators that
have zero expectation value in the $q\bar{q}$ color singlet.

We wish to derive an effective Hamiltonian that  acts at the hadronic scale
by lowering the similarity cutoff perturbatively
as low as possible, hopefully down
 to the scale
at which the bound state is well approximated by its two particle component,
$\bar{q} q$. This may not be possible for all systems. The coupling may
become too large for perturbation theory to be reliable before the
higher Fock components are eliminated. We know, however,
 that it is possible in QED,
and we believe that the  success of the  CQM suggests the same for QCD.
The most favorable systems as far as the $q\bar{q}$ and nonrelativistic
 approximations are
concerned are heavy quarkonia. However heavy mesons containing a light quark
provide a better tool for  testing our approach qualitatively, because they
have fine structure that allows us to separately test spin and orbital
angular momentum-dependent operators in the Hamiltonian, in the regime
which is  dominated by the confining interaction. This means that the
scaling of momenta and energies is determined by the confining interaction.
As the Coulomb part of the potential becomes more important, the simple
scaling analysis breaks down, until the Coulomb potential dominates over the
confining potential. Then the momenta and energies scale as in QED.
We expect heavy quarkonia to be in the mixed regime, so the simple scaling
analysis is probably not going to be reliable.
Study of the spin and orbital angular momentum-dependent operators, which
are generated by  the similarity transformation to second order in the
coupling, will be done in later work.

 In heavy-light systems there is only one heavy quark, but we do not
 yet know whether the  light degrees of freedom can be approximated by just one
constituent in our approach. Further,
as we show below, heavy-light mesons are qualitatively different from
heavy quarkonia, but there are similarities with other
mesonic systems:  strange  mesons and isospin 1 light mesons.
So some of what we learn from heavy mesons may be generalized to light mesons.
We will choose B mesons  to check whether
we can fit spectra with reasonable parameters.

After bringing the cutoff down to the hadronic scale, we do bound state
calculations. We have to decide which terms will be treated as the dominant
interactions
and put in  the unperturbed Hamiltonian, and which ones will be
treated as perturbations.

In order to gain qualitative insight it is useful to study the nonrelativistic
limit and to rewrite the bound state equation in  position space.
It turns out that in the nonrelativistic limit light-front dynamics
naturally reduce to equal-time dynamics, which implies that angular momenta
become kinematically defined. It is possible to transform
light-front coordinates to  equal-time coordinates by a specific change of
variables \cite{coester} without taking the nonrelativistic limit. However,
the point we want to make here is that in the nonrelativistic limit
equal-time  dynamics arise naturally. This will become  clearer below.

Nonrelativistic reduction can be justified at best only for the lowest lying
states. We may need to do a series of bound state calculations with small
coupling, and then extrapolate to the physical value of the strong coupling
if this is large \cite{thelongpaper}.
However, we are not yet at the stage where we can
carry out the strategy with confidence that we have complete
 control over the bound state perturbation theory. The similarity
transformation generates effective operators.
The primary motivation of this work is to initiate the study of these
operators. Further, the issue of chiral symmetry breaking
terms, that is, what operators  have to be
added to the Hamiltonian by hand  to restore the effects of
zero modes removed by a small longitudinal cutoff, is not resolved yet.
Nonrelativistic reduction provides a framework in which these and other
questions can be studied.

The main questions answered in this calculation are:
\begin{enumerate}
\item{Two-body potentials generated by the similarity transformation
include the Coulomb potential
 and a logarithmic confining potential already
at order $g^2$. Does this effective Hamiltonian contain enough structure
to provide a starting point for studying bound states in QCD?
For the spin-independent part of the effective Hamiltonian, which we study
here, the answer is: Yes.}
\item{ The similarity transformation
generates new effective interactions. Is there a
simple way to see how these new operators affect spectra?
Yes.}
\item{ In order to answer the previous questions,  one has to make some
simplifications of the original Hamiltonian, including but not restricted to
nonrelativistic reduction (for details see section 3).
We do not focus on quantitative analysis, but we want to ask whether  the
calculation is at all
reasonable; that is, can we fit any data with a set of reasonable parameters?
Yes.}
\end{enumerate}

The paper is organized as follows. In the second section we briefly review
the general strategy, sketching the similarity transformation and coupling
coherence. The third section gives our calculation - $q\bar{q}$ to
order $g^2$. First, we find the effective Hamiltonian using the similarity
transformation and coupling coherence. Then we split the Hamiltonian into a
part
which is treated nonperturbatively in the bound state calculation and a part
which is treated perturbatively.
In this paper, we solve the leading order
problem, and show that it  reduces to a simple
Schr{\H{o}}dinger equation in the nonrelativistic limit. We show a simple way
to qualitatively analyze the physics behind this Schr{\H{o}}dinger equation.
Then we present the numerical results. The last section contains our summary
and
conclusions.

\section{Two steps to solving QCD - general strategy}

In this section we  briefly review the general strategy, first outlined in
ref. \cite{thelongpaper}.
We start with the canonical light-front QCD Hamiltonian in light-cone gauge,
 $A^+_a=0$. We
will not explicitly show terms that are not important for the specific
calculations presented in the
next section. For a detailed discussion of the light-front
Hamiltonian see  \cite{thelongpaper,xy}
and references therein.
Ignoring purely gluonic terms that do not affect the effective $q\bar{q}$
Hamiltonian until fourth order in $g$,
\begin{eqnarray}
H= H_{\rm free} + V_1 +V_2   \   \  ;
\end{eqnarray}
where $H_{\rm free}$ is the free light-front Hamiltonian:
\begin{eqnarray}
H_{free} & = &
\sum_f\int {d^3p\over{(2\pi)^3 2 p^+}} {p^{\perp 2}+m_f^2\over{p^+}}
\left(b_f^{\dagger}b_f + d_f^{\dagger}d_f\right)  \  \  .
\end{eqnarray}
There is a sum over flavors,
$m_f$ is a quark  mass,
$b_f, d_f$ are quark and antiquark
annihilation operators;
\begin{eqnarray}
V_1= g \int dx^- d^2 x_{\perp} \bar{\psi}
{\not{\hbox{\kern-4pt $A$}}} \psi
\end{eqnarray}
contains the standard order $g$ quark-gluon coupling. Here $\psi$ and
 $A^{\mu} \equiv \sum_{a} A_a^{\mu} T^a$ are free
light-front fields:
\begin{eqnarray*}
\psi =\left( \begin{array}{c} \psi_+ \\
\psi_- =
{1\over{\partial ^+}}(-i\vec{\alpha}^{\perp}\cdot \vec{\partial}_{\perp}
+\beta m) \psi_+ \end{array} \right)
\end{eqnarray*}
and
\begin{eqnarray*}
A^{\mu} = \left( A^+=0,\  A^-={2\over{\partial ^+}}\vec{\partial}^{\perp} \cdot
\vec{A}^{\perp}, \  \vec{A}^{\perp} \right).
\end{eqnarray*}
The constrained fields, $\psi _-$ and $A^-$, are replaced by
functions of the physical degrees of freedom resulting in new terms in the
canonical Hamiltonian, among which
\begin{eqnarray}
V_2 =  - 2g^2 \int dx^- d^2 x_\perp
 (\psi_+^{\dagger} T^a \psi_+ ) \left( \frac{1}
{\partial^+} \right)^2 (\psi_+^{\dagger} T^a \psi_+)
\end{eqnarray}
is the instantaneous gluon exchange between two fermions.

We regulate the Hamiltonian with cutoffs on the change in free energy at each
interaction vertex and a cutoff on longitudinal momentum fractions that is
taken to zero at the end of the calculation. Then
 we use a  similarity transformation
The similarity transformations
form a renormalization group. Repeated transformations generate a sequence of
Hamiltonians with decreasing cutoff.
A Hamiltonian in the sequence
is related to the previous one by:
\begin{eqnarray}
H_{\Lambda_{n-1}} = U^{-1}(\Lambda_{n-1}) \ H_{\Lambda_{n}} U(\Lambda_{n-1})
 \  \  .
\end{eqnarray}
 $U $ is a unitary matrix and can be written as $U=e^{iR}$, where $R$ is
hermitian and has
an expansion in powers of
the nondiagonal part of the Hamiltonian. We have chosen $n$ so
that it decreases as the cutoff decreases.

The similarity
transformation is designed
to bring the cutoff Hamiltonian to a band diagonal
form, while avoiding problems of small energy denominators in perturbative
expansions \cite{similarity}.
In particular, the transformed Hamiltonian is required to be band diagonal
relative to the new scale. This means that the matrix elements between states
that differ in energy by more  than the new cutoff must be zero
for the simple step function cutoffs we employ here.
This requirement has implications for the matrix elements of  $R$.
Given $R$, one can find the transformed Hamiltonian.

The initial cutoff
%similarity transformation
destroys manifest symmetries and one of the criteria for the renormalized
Hamiltonian is that it restores the symmetries,
albeit not necessarily in  manifest form.
If the similarity transformation can be done analytically, it is possible to
use coupling coherence \cite{pertren,couplcoh}
to completely fix
the renormalized Hamiltonian without explicit reference to symmetries.
 The basic idea of coupling coherence is that in the Hamiltonian
restricted by symmetries, the
strengths of all operators are not independent but depend only on a finite
number of independent canonical
parameters; so that under a transformation, the Hamiltonian
reproduces itself in form exactly, apart from the change of the
explicit cutoff and
the running of  those  few independent couplings.
All dependence on the
cutoff is absorbed into the
independent running couplings.
Once one obtains a Hamiltonian that  reproduces itself as the cutoff is
lowered, any initial cutoff can be sent to infinity.

%*COPIED FROM THE PERRY POLAND
In order to use coupling coherence we need to study how the Hamiltonian
changes when the cutoff changes.
Let $H_{\Lambda_n}=H_{\rm free}+v$, where $H_{\rm free}$ is a free
Hamiltonian and
 $v \equiv V_1 +V_2$
is cut off so that
\begin{equation}
\langle \phi_i | v | \phi_j \rangle = 0 \;,
\end{equation}

\noindent if $|E_{0i}-E_{0j}| > {\Lambda_n^2\over{{\cal P}^+}}$;
where $H_{\rm free} |\phi_i\rangle =
E_{0i} |\phi_i \rangle$.
The similarity cutoff, ${\Lambda_n ^2 \over{{\cal P}^+}}$ with the
dimension of light-front energy, consists of
$\Lambda_n ^2$, which carries a dimension of transverse mass squared, and an
 arbitrary longitudinal momentum reference scale ${\cal P}^+$.
If this cutoff is lowered to
 ${\Lambda_{n-1}^2\over{{\cal P}^+}}$ by the similarity transformation, the
new Hamiltonian matrix elements to ${\cal O}(v^2)$ are \cite{robpoland}:
\begin{eqnarray}
\lefteqn{ H_{\Lambda_{n-1} ab} =
\langle \phi_a|\  H_{\rm free}+v \ |\phi_b\rangle }\nonumber \\
& - \sum_k v_{ak} v_{kb} \Biggl[
{\theta\bigl(|\Delta_{ak}|-{\Lambda_{n-1}^2\over{{\cal P}^+}} \bigr)
\theta\bigl(|\Delta_{ak}|-|\Delta_{bk}|\bigr) \over E_{0k}-E_{0a} }
+ {\theta\bigl(|\Delta_{bk}|-{\Lambda_{n-1}^2\over{{\cal P}^+}}\bigr)
\theta\bigl(|\Delta_{bk}|-|\Delta_{ak}|\bigr) \over E_{0k}-E_{0b} }
\Biggr],
\end{eqnarray}
\noindent where $\Delta_{ij}=E_{0i}-E_{0j}$ and
$|E_{0a}-E_{0b}|<{\Lambda_{n-1}^2\over{{\cal P}^+}}$, and
 there are implicit cutoffs in this expression because
the matrix elements of $v$ have already been cut off so that
$v_{ij}=0$ if $|E_{0i}-E_{0j}|>{\Lambda_{n}^2\over{{\cal P}^+}}$.

To this order, a coupling coherent
Hamiltonian reproduces itself, with the only change being
$\Lambda_{n} \rightarrow \Lambda_{n-1}$.
At the third order one begins to see the quark-gluon coupling run.
If the Hamiltonian reproduces itself, the index $n$ becomes irrelevant.
The solution is found by noting that
we need the partial sum above to be added to an interaction in $v$
that is expressed as a sum, so that the transformation merely changes
the limits on the sum in a simple fashion (for details see \cite{P2}).
There are two
possibilities.  The first is:
\begin{eqnarray}
\lefteqn{ H_{ab} = \langle \phi_a|h_0+v|\phi_b\rangle  } \nonumber\\
&  -\sum_k v_{ak} v_{kb} \left[
{\theta\bigl(|\Delta_{ak}|-{\Lambda_{n-1}^2\over{{\cal P}^+}} \bigr)
\theta\left(|\Delta_{ak}|-|\Delta_{bk}|\right) \over{ E_{0k}-E_{0a}} }
+ {\theta\bigl(|\Delta_{bk}|-{\Lambda_{n-1}^2\over{{\cal P}^+}}\bigr)
\theta\left(|\Delta_{bk}|-|\Delta_{ak}|\right) \over E_{0k}-E_{0b}}
\right] ,
\end{eqnarray}
and the second is:
\begin{eqnarray}
\lefteqn{ H_{ab} = \langle \phi_a|h_0+v|\phi_b\rangle  } \nonumber\\
& +  \sum_k  v_{ak} v_{kb} \Biggl[
{\theta\bigl({\Lambda_{n-1}^2\over{{\cal P}^+}}-|\Delta_{ak}| \bigr)
\theta\bigl(|\Delta_{ak}|-|\Delta_{bk}|\bigr) \over E_{0k}-E_{0a} }
+ {\theta\bigl({\Lambda_{n-1}^2\over{{\cal P}^+}}-|\Delta_{bk}| \bigr)
\theta\bigl(|\Delta_{bk}|-|\Delta_{ak}|\bigr) \over E_{0k}-E_{0b} }
\Biggr] .
\end{eqnarray}

Note that $v$ in these expressions is the same as that
above only to first order.  The coupling coherent interaction in $H$
is written as a power series in $v$ which reproduces itself under the
transformation, except the cutoff changes.  In higher orders the
canonical couplings  also run.
To decide which of
these equations to use one must in principle go to higher orders, but
in practice it is usually obvious which choice is correct.  For example,
the first solution provides effective two-body interactions from one-gluon
exchange; while the second provides the relevant part of the quark self-
energy. If chosen otherwise, the new terms would make the effective
Hamiltonian divergent.

At the end of this first step, the Hamiltonian is renormalized and the scale
relative to which it is band diagonal is a hadronic scale.{\footnote{A
specific value of the ``hadronic scale'' is to be specified by fitting
spectra.}} The
effective Hamiltonian contains complicated potentials, which result from
eliminating the coupling between high and low energy states. It still contains
emission and absorption interactions, but these no longer mix states of high
and
low energies.
%\textfraction{0.1}

In the second step, the Hamiltonian which is now given as an expansions in $g$
is regrouped from the point of view of what is important in the bound state.
 Based on the success of the CQM, we believe that emission and absorption
processes, which would mix different Fock states, are suppressed with respect
to interactions that do not change particle number
for low-lying states and sufficiently low cutoff. Thus the particle number
changing interactions
will be treated perturbatively using bound state perturbation theory.
This is clearly justified if the gluons are
massive \cite{thelongpaper},
 but even for massless gluons one can argue that the interacting gluon
effectively acquires mass related to the confining scale \cite{P2}. It is
therefore plausible to assume that adding an extra constituent is suppressed
even in this case.

Next, we use constituent masses.
It is possible to add the constituent mass at zeroth order and subtract it
at a higher order in bound state perturbation theory, as outlined in
 \cite{thelongpaper}.  This issue starts to be important
 when the effective Hamiltonian is calculated to higher orders
 and  one tries
to see that the approximations in the leading
 order bound state calculation
 lead to convergent results.

In the calculation presented here we also use a
nonrelativistic reduction and we choose a rotationally symmetric $H_0$.
We want to minimize higher order corrections, but we do not yet know what
restores the
rotational symmetry (e.g., adding extra gluons or terms of higher order in
$g$). The answer to this question may change our choice of $H_0$ in future
calculations. At least for now, we choose the confining potential
 such that it does not yield any corrections in first order bound state
perturbation theory.
The corrections which come from the rotationally
noninvariant terms are not suppressed by powers of the coupling, despite the
fact that they  enter at the second order of  bound state perturbation theory.
This is  the same order at which one has to include
  emission and absorption processes, unless those are suppressed
nonperturbatively (e.g. if the gluons are massive).
Therefore, it is not clear whether it is meaningful to consider the
corrections due to
rotationally noninvariant terms without including $q \bar{q}g$.
Nevertheless, we evaluated these corrections for the ground state and they
are of order  a few percent in a region where the confining potential
dominates over the Coulomb potential.

\section{The simple QCD calculation - $q{\bar{q}}$ to order $g^2$}

In this section we  discuss  mesons, i.e. color singlet QCD bound states
whose valence constituents are a quark and an antiquark. The
masses of the constituents are arbitrary but nonzero.
We expect that the qualitative aspects of the study are relevant to
all $q\bar{q}$ systems with a possible exception of light isospin zero
mesons{\footnote{For light mesons with I=0 we expect
that operators of O$(g^4)$ will
 play an important role.}. We  fit  B mesons.
%\textfraction{.1}
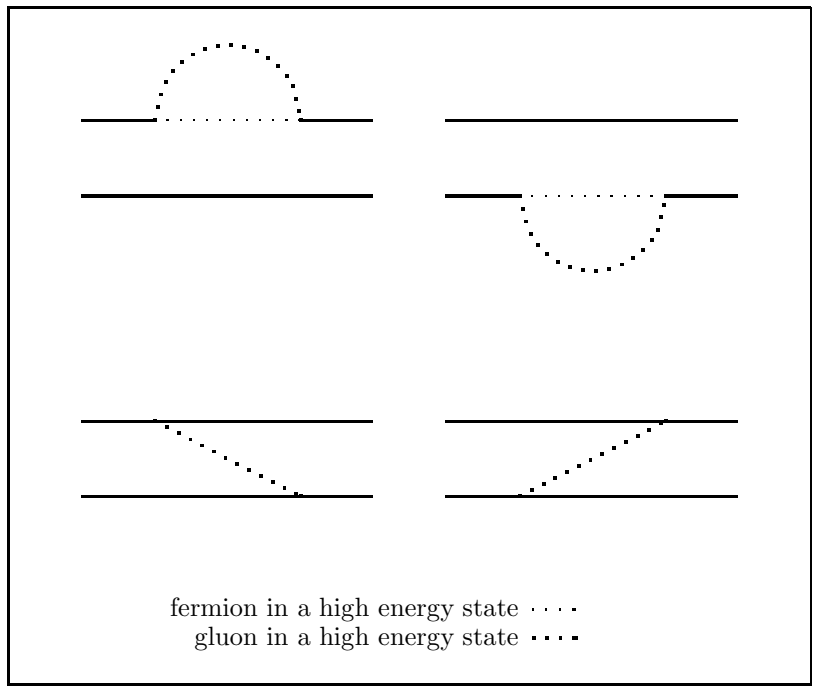
\begin{figure}[th!]
\begin{center}
% GNUPLOT: LaTeX picture
\setlength{\unitlength}{0.240900pt}
\ifx\plotpoint\undefined\newsavebox{\plotpoint}\fi
\sbox{\plotpoint}{\rule[-0.200pt]{0.400pt}{0.400pt}}%
\begin{picture}(1500,1200)(0,0)
\font\gnuplot=cmr10 at 10pt
\gnuplot
\sbox{\plotpoint}{\rule[-0.200pt]{0.400pt}{0.400pt}}%
\put(176.0,68.0){\rule[-0.200pt]{303.534pt}{0.400pt}}
\put(1436.0,68.0){\rule[-0.200pt]{0.400pt}{256.318pt}}
\put(176.0,1132.0){\rule[-0.200pt]{303.534pt}{0.400pt}}
\put(806,1177){\makebox(0,0){ }}
\put(176.0,68.0){\rule[-0.200pt]{0.400pt}{256.318pt}}
\sbox{\plotpoint}{\rule[-0.400pt]{0.800pt}{0.800pt}}%
\put(291,836){\usebox{\plotpoint}}
\put(291.0,836.0){\rule[-0.400pt]{110.332pt}{0.800pt}}
\put(291,955){\usebox{\plotpoint}}
\put(291.0,955.0){\rule[-0.400pt]{27.463pt}{0.800pt}}
\put(634,955){\usebox{\plotpoint}}
\put(634.0,955.0){\rule[-0.400pt]{27.703pt}{0.800pt}}
\sbox{\plotpoint}{\rule[-0.200pt]{0.400pt}{0.400pt}}%
\put(405,955){\usebox{\plotpoint}}
\multiput(405,955)(20.756,0.000){12}{\usebox{\plotpoint}}
\put(634,955){\usebox{\plotpoint}}
\sbox{\plotpoint}{\rule[-0.500pt]{1.000pt}{1.000pt}}%
\put(405,955){\usebox{\plotpoint}}
\multiput(405,955)(4.858,20.179){3}{\usebox{\plotpoint}}
\put(421.66,1014.64){\usebox{\plotpoint}}
\put(433.36,1031.75){\usebox{\plotpoint}}
\put(447.71,1046.62){\usebox{\plotpoint}}
\put(464.77,1058.40){\usebox{\plotpoint}}
\multiput(469,1061)(19.159,7.983){0}{\usebox{\plotpoint}}
\put(483.66,1066.82){\usebox{\plotpoint}}
\put(503.83,1071.51){\usebox{\plotpoint}}
\multiput(507,1072)(20.694,1.592){0}{\usebox{\plotpoint}}
\put(524.49,1072.63){\usebox{\plotpoint}}
\multiput(532,1072)(20.514,-3.156){0}{\usebox{\plotpoint}}
\put(545.06,1069.98){\usebox{\plotpoint}}
\put(564.74,1063.41){\usebox{\plotpoint}}
\put(582.69,1053.21){\usebox{\plotpoint}}
\multiput(583,1053)(16.451,-12.655){0}{\usebox{\plotpoint}}
\put(598.71,1040.08){\usebox{\plotpoint}}
\put(611.90,1024.17){\usebox{\plotpoint}}
\multiput(621,1009)(4.858,-20.179){3}{\usebox{\plotpoint}}
\put(634,955){\usebox{\plotpoint}}
\sbox{\plotpoint}{\rule[-0.400pt]{0.800pt}{0.800pt}}%
\put(863,836){\usebox{\plotpoint}}
\put(863.0,836.0){\rule[-0.400pt]{27.703pt}{0.800pt}}
\put(863,955){\usebox{\plotpoint}}
\put(863.0,955.0){\rule[-0.400pt]{110.332pt}{0.800pt}}
\put(1207,836){\usebox{\plotpoint}}
\put(1207.0,836.0){\rule[-0.400pt]{27.463pt}{0.800pt}}
\sbox{\plotpoint}{\rule[-0.200pt]{0.400pt}{0.400pt}}%
\put(978,186){\makebox(0,0)[r]{fermion in a high energy state}}
\multiput(1000,186)(20.756,0.000){4}{\usebox{\plotpoint}}
\put(1066,186){\usebox{\plotpoint}}
\put(978,836){\usebox{\plotpoint}}
\multiput(978,836)(20.756,0.000){12}{\usebox{\plotpoint}}
\put(1207,836){\usebox{\plotpoint}}
\sbox{\plotpoint}{\rule[-0.500pt]{1.000pt}{1.000pt}}%
\put(978,836){\usebox{\plotpoint}}
\multiput(978,836)(4.858,-20.179){3}{\usebox{\plotpoint}}
\put(994.46,776.23){\usebox{\plotpoint}}
\put(1005.83,758.95){\usebox{\plotpoint}}
\put(1020.60,744.46){\usebox{\plotpoint}}
\put(1037.78,732.88){\usebox{\plotpoint}}
\multiput(1041,731)(18.845,-8.698){0}{\usebox{\plotpoint}}
\put(1056.59,724.20){\usebox{\plotpoint}}
\put(1076.74,719.50){\usebox{\plotpoint}}
\multiput(1080,719)(20.684,-1.724){0}{\usebox{\plotpoint}}
\put(1097.40,718.42){\usebox{\plotpoint}}
\put(1117.98,721.00){\usebox{\plotpoint}}
\multiput(1118,721)(19.838,6.104){0}{\usebox{\plotpoint}}
\put(1137.39,728.19){\usebox{\plotpoint}}
\put(1155.75,737.86){\usebox{\plotpoint}}
\multiput(1156,738)(16.451,12.655){0}{\usebox{\plotpoint}}
\put(1171.65,751.09){\usebox{\plotpoint}}
\put(1184.48,767.35){\usebox{\plotpoint}}
\multiput(1194,782)(4.858,20.179){3}{\usebox{\plotpoint}}
\put(1207,836){\usebox{\plotpoint}}
\sbox{\plotpoint}{\rule[-0.400pt]{0.800pt}{0.800pt}}%
\put(291,482){\usebox{\plotpoint}}
\put(291.0,482.0){\rule[-0.400pt]{110.332pt}{0.800pt}}
\put(291,364){\usebox{\plotpoint}}
\put(291.0,364.0){\rule[-0.400pt]{110.332pt}{0.800pt}}
\sbox{\plotpoint}{\rule[-0.500pt]{1.000pt}{1.000pt}}%
\put(405,482){\usebox{\plotpoint}}
\multiput(405,482)(18.450,-9.507){13}{\usebox{\plotpoint}}
\put(634,364){\usebox{\plotpoint}}
\sbox{\plotpoint}{\rule[-0.400pt]{0.800pt}{0.800pt}}%
\put(863,482){\usebox{\plotpoint}}
\put(863.0,482.0){\rule[-0.400pt]{110.332pt}{0.800pt}}
\put(863,364){\usebox{\plotpoint}}
\put(863.0,364.0){\rule[-0.400pt]{110.332pt}{0.800pt}}
\sbox{\plotpoint}{\rule[-0.500pt]{1.000pt}{1.000pt}}%
\put(978,141){\makebox(0,0)[r]{gluon in a high energy state}}
\multiput(1000,141)(20.756,0.000){4}{\usebox{\plotpoint}}
\put(1066,141){\usebox{\plotpoint}}
\put(978,364){\usebox{\plotpoint}}
\multiput(978,364)(18.450,9.507){13}{\usebox{\plotpoint}}
\put(1207,482){\usebox{\plotpoint}}
\end{picture}
\end{center}
\caption{Diagrams representing similarity transformation to order $g^2$ for
$q\bar{q}$. The top line represents new effective one-body operators, the
bottom line are two-body operators. We use untypical lines for gluons and
fermions to emphasize that these represent new operators, not Feynman
diagrams.}
%Figure 1.
\end{figure}

To order $g^2$,  the similarity transformation
is represented by a few diagrams, as shown in figure 1,
so it is possible to find the effective Hamiltonian analytically.
Let us work out in detail one of the operators, and then list the
results for remaining ones.

Let us consider matrix elements involving one gluon exchange between a quark
and
an antiquark (see the bottom two diagrams in figure 1).
If the Hamiltonian is band-diagonal relative to a
scale
${{\Lambda_n^2}\over{{\cal P}^+}}$, then only those matrix elements in which
the
energy difference is less than this value are nonzero:
\begin{eqnarray}
   & -g_{\Lambda_{n}}^2
   \bar{u}(p_2, \sigma _2) \gamma ^{\mu} u(p_1, \sigma _1)
\bar{v}(k_2, \lambda_2) \gamma ^{\nu} v(k_1, \lambda _1)
\langle T_a T_b \rangle
\nonumber\\
  & \left[
{\theta(q^+)\over{q^+}}
D_{\mu \nu}(q)
\left({\theta({\Lambda_n^2 \over{{\cal P}^+}} -\vert D_1\vert)
\theta(\vert D_1\vert -\vert D_2\vert )\over{D_1}}
+ {\theta({\Lambda_n^2 \over{{\cal P}^+}} -\vert D_2\vert)
\theta(\vert D_2\vert -\vert D_1\vert )\over{D_2}}\right)
\right.
\nonumber\\
  & \left.
 \  \  +
 {\theta(-q^+)\over{-q^+}}
 D_{\mu \nu}(-q)
\left({\theta({\Lambda_n^2 \over{{\cal P}^+}} -\vert -D_1\vert)
\theta(\vert D_1\vert -\vert D_2\vert ) \over{-D_1}}
 + {\theta({\Lambda_n^2 \over{{\cal P}^+}} -\vert -D_2\vert)
\theta(\vert D_2\vert -\vert D_1\vert )\over{-D_2}}\right)
\right] ,
\end{eqnarray}
%\vfill
%\pagebreak
\noindent where $p_i$, $k_i$ are light-front three-momenta carried
by the constituents;
$\sigma _i$, $\lambda_i$ are their light-front helicities;
$u(p, \sigma )$, $v(k, \lambda )$ are their spinors \cite{lebro};
index $i=1,2$ refers to the initial and final states, respectively;
$D_{\mu \nu}(q) = {{q^{\perp}}^2\over{{q^+}^2}}\eta_{\mu}\eta_{\nu}
 + {1\over{q^+}}
\left(\eta_{\mu}{q^{\perp}}_{\nu} + \eta_{\nu}{q^{\perp}}_{\mu}\right)
- g^{\perp}_{\mu \nu}$ is the gluon propagator in light-front gauge
\cite{bass},
%\begin{eqnarray*}
%$D_{\mu \nu}(q) = {{q^{\perp}}^2\over{{q^+}^2}}\eta_{\mu}\eta_{\nu}
% + {1\over{q^+}}
%\left(\eta_{\mu}{q^{\perp}}_{\nu} + \eta_{\nu}{q^{\perp}}_{\mu}\right)
%- g^{\perp}_{\mu \nu},$
%\end{eqnarray*} \noindent
%where
$\eta _\mu = ( 0, \eta _+ = 1,0,0)$;
$\vec{q} = \vec{p}_1 - \vec{p}_2$ is the exchanged momentum,
$q^- ={ {q^{\perp}}^2\over{q^+}}$;
$D_1$, $D_2$ are energy denominators:
$D_1 = p_1^- -p_2^- -q^-$ and $D_2 = k_2^- -k_1^- - q^-$.

It is convenient to use Jacobi momenta. Setting the total transverse
momentum to be zero, the momenta of the constituents  are:
\begin{eqnarray*}
p_i^+  =  x_i P^+ &, & p_i^{\perp}  =  \kappa _i ^{\perp} \  \  ; \nonumber\\
k_i^+  =  (1-x_i) P^+ &, & k_i^{\perp}  =  - \kappa _i ^{\perp}  \  \  .
\end{eqnarray*}
Let the mass of the constituent with momentum $p$ be $m_a$ and the mass of the
other constituent be $m_b$.
The denominators in terms of Jacobi momenta are:
\begin{eqnarray}
D_1 = {1\over{P^+}}\left({{\kappa_1^{\perp}}^2 +m_a^2\over{x_1}}
-{{\kappa_2^{\perp}}^2 +m_a^2\over{x_2 }}-
{(\kappa_1 ^{\perp} -\kappa_2^{\perp})^2\over{x_1 -x_2}} \right),\nonumber\\
D_2 = {1\over{P^+}}\left({{\kappa_2^{\perp}}^2 +m_b^2\over{1-x_2}}
-{{\kappa_1^{\perp}}^2 +m_b^2\over{1-x_1 }}-
{(\kappa_1 ^{\perp} -\kappa_2^{\perp})^2\over{x_1 -x_2}} \right).
\end{eqnarray}

When the scale is lowered to ${\Lambda^2_{n-1}\over{{\cal P}^+}}$ by the
similarity transformation, all matrix elements
in which the energy jump is larger than the new cutoff are zeroed. The
effects of couplings which are removed have to be put directly in the
Hamiltonian as new effective interactions. In this case, the new effective
interactions according to $(8)$ are:
\begin{eqnarray}
  & -g_{\Lambda_{n-1}}^2
  \bar{u}(p_2, \sigma _2) \gamma ^{\mu} u(p_1, \sigma _1)
\bar{v}(k_2, \lambda_2) \gamma ^{\nu} v(k_1, \lambda _1)
\langle T_a T_b \rangle
\nonumber\\
  & \left[
 {\theta(q^+)\over{q^+}}
D_{\mu \nu}(q)
\left({\theta(\vert D_1\vert-{\Lambda_{n-1}^2 \over{{\cal P}^+}})
\theta(\vert D_1\vert -\vert D_2\vert )\over{D_1}}
+ {\theta(\vert D_2\vert-{\Lambda_{n-1}^2 \over{{\cal P}^+}})
\theta(\vert D_2\vert -\vert D_1\vert )\over{D_2}}\right)
\right.
\nonumber\\
  & \  \  +
 {\theta(-q^+)\over{-q^+}}
\left.
D_{\mu \nu}(-q)
\left({\theta(\vert -D_1\vert-{\Lambda_{n-1}^2 \over{{\cal P}^+}})
\theta(\vert D_1\vert -\vert D_2\vert ) \over{-D_1}}
 + {\theta(\vert -D_2\vert-{\Lambda_{n-1}^2 \over{{\cal P}^+}})
\theta(\vert D_2\vert -\vert D_1\vert )\over{-D_2}}\right)
\right] .
\end{eqnarray}

This will  repeat as the cutoff is lowered. Once the interaction
reproduces
itself in form
the initial cutoff can be sent to
infinity and we can lower the cutoff to the scale of interest.
However, as  the cutoff decreases, the coupling increases
and at some point it becomes invalid to use perturbation theory
to further lower the cutoff scale.
The interaction in the effective Hamiltonian at the hadronic scale $\Lambda$ is
thus:
\begin{eqnarray}
  & -g_{\Lambda }^2 \bar{u}(p_2, \sigma _2) \gamma ^{\mu} u(p_1, \sigma _1)
\bar{v}(k_2, \lambda_2) \gamma ^{\nu} v(k_1, \lambda _1)
\langle T_a T_b \rangle
\nonumber\\
 & \left[
 {1\over{q^+}}
D_{\mu \nu}(q)
\left({\theta(\vert D_1\vert-{\Lambda^2 \over{{\cal P}^+}})
\theta(\vert D_1\vert -\vert D_2\vert )\over{D_1}}
+ {\theta(\vert D_2\vert-{\Lambda^2 \over{{\cal P}^+}})
\theta(\vert D_2\vert -\vert D_1\vert )\over{D_2}}\right)
\right] .
\end{eqnarray}
\noindent Here we summed the two terms corresponding to the two
time-ordered diagrams in figure~1.

Similarly, one can find effective one-body operators (self-energies)(see
 the top two diagrams in figure 1):
\begin{eqnarray}
{\alpha_{\Lambda} C_F \over{2\pi P^+}} \left\{
2 {P^+\over{{\cal P}^+}}\Lambda^2 \log\left({P^+\over{\epsilon {\cal P}^+}}
\right)
+ 2 {P^+\over{{\cal P}^+}}\Lambda^2
\log {x^2 {P^+\over{{\cal P}^+}}\Lambda^2 \over{x{P^+\over{{\cal P}^+}}
\Lambda^2 +m^2  }} \right. \nonumber\\
- {3\over{2}}{P^+\over{{\cal P}^+}}\Lambda^2
+{1\over{2}}{m^2 {P^+\over{{\cal P}^+}}\Lambda^2
\over{x{P^+\over{{\cal P}^+}}\Lambda^2 +m^2  }} \nonumber\\
 \left.
+ 3 {m^2\over{x}}\log
 {m^2\over{x {P^+\over{{\cal P}^+}} \Lambda^2 +m^2  }} \right\} \  \  .
\end{eqnarray}
Here $m$ and $x$ stands for either $m_a$ and $x_a$, or $m_b$ and $x_b$.
The first term is infrared divergent ($\epsilon$ is an infinitesimal cutoff on
a longitudinal momentum
 taken to zero at the end of calculation) but it exactly cancels with
the infrared divergence in the effective two-body operator $(13)$, if the
$q\bar{q}$ pair
is in a color singlet \cite{P2}.

By finding these counterterms we have
completed the first step of the calculation.
Let us summarize the effective Hamiltonian:
\begin{eqnarray}
H_{\rm eff} = H_{\rm free} + V_1 +  V_{2} +  V_{2 \ {\rm eff}} \  \  ,
\end{eqnarray}
where $H_{\rm free}$ is the kinetic energy; $V_1$ is ${\cal O}(g)$
emission and absorption with
nonzero matrix elements
only between states with energy difference smaller than the
hadronic scale ${\Lambda^2\over{{\cal P}^+}}$; $V_{2} $ is
${\cal O}(g^2)$
instantaneous interaction (with no cutoff) and $V_{2 \ {\rm eff}}$ includes
the effective
interactions, also ${\cal O}(g^2)$, given in previous formulae.

This brings us to the second step: we have to regroup and
approximate the Hamiltonian for the purpose of bound state calculations.
As mentioned above, the emission and absorption are not  included in $H_0$.
We  include kinetic energy, instantaneous fermion exchange,
 self-energies and the most infrared divergent
piece of the effective interaction arising from one gluon exchange. Even
this is still quite complicated as a starting point to
gain intuition, so we
consider a nonrelativistic limit of this Hamiltonian.
As before, we  derive in detail results for a specific operator and list
the results for the remaining ones.

Let us consider the most infrared divergent piece of the effective
interaction arising from one
gluon exchange in a color singlet:
\begin{eqnarray}
 & -g_{\Lambda}^2 C_F \bar{u}(p_2, \sigma _2) \gamma ^{+} u(p_1, \sigma _1)
\bar{v}(k_2, \lambda_2) \gamma ^{+} v(k_1, \lambda _1)
\nonumber\\
  &  \left[
 {1\over{q^+}}
 { {q^{\perp}}^2\over{ {q^+}^2}}
\left({\theta(\vert D_1\vert-{\Lambda^2 \over{{\cal P}^+}})
\theta(\vert D_1\vert -\vert D_2\vert )\over{D_1}}
+ {\theta(\vert D_2\vert-{\Lambda^2 \over{{\cal P}^+}})
\theta(\vert D_2\vert -\vert D_1\vert )\over{D_2}}\right)
\right] .
\end{eqnarray}

Let us denote
$M_{ab} \equiv m_a +m_b$,
introduce a new variable for the longitudinal momentum fraction
 $\eta$ such that $x_a= {m_a\over{M_{ab}}} - \eta$, $x_b=1-x_a$ and then
 make an  expansion in powers of~$\eta$.

To the lowest order in momenta,
both energy denominators
reduce to
\begin{eqnarray}
D_1 = D_2 = - {1\over{P^+}}\left( M_{ab}^2 (\eta_1 -\eta_2) +
 {(\kappa_1^{\perp} - \kappa_2^{\perp})^2\over{(\eta_1 -\eta_2) }}\right) .
\end{eqnarray}
Here $(\kappa_1^{\perp} - \kappa_2^{\perp})^2 = q^{\perp 2}$. If we further
identify $q_z^2 \equiv M_{ab}^2 (\eta_1 - \eta_2)^2$,
then  the expression $(16)$ reduces to:
\begin{eqnarray}
4 g_{\Lambda}^2 \sqrt{x_1(1-x_1)x_2 (1-x_2)} M_{ab}^2
{ q^{\perp 2}\over{q_z^2 \ \vec{q}^2}} (1-\theta_{below}) \  \  ,
\end{eqnarray}
where
$$\theta_{below} \equiv \theta \left({\Lambda^2 \over{{\cal P}^+}}-
{ M_{ab} \vec{q}^2\over{P^+ \vert q_z
\vert}}\right).$$
{}From now on, we  drop the omnipresent $\sqrt{x_1(1-x_1)x_2 (1-x_2)}$,
because it cancels exactly with the same factor in the definition of the
wavefunction (see eqn. $(30)$ in the next section).

Adding this to a canonical term of order $g^2$
which has a similar structure,
namely the instantaneous interaction:
\begin{eqnarray}
 - 4g_{\Lambda}^2 C_F
{1\over{(\eta_1 -\eta_2)^2}}
   =  - 4g_{\Lambda }^2C_F
{M_{ab}^2\over{q_z^2}} \  \  ,
\end{eqnarray}
leads to the following interaction:
\begin{eqnarray}
-{4g^2 C_F M_{ab}^2 \over{\vec{q}^2}} -
  {4g^2 C_F M_{ab}^2 q^{\perp 2}\over{ q_z^2 \vec{q}^2}}
\theta_{below} \   \   .
\end{eqnarray}
The first term is the Coulomb potential.
The scale dependent part of the
interaction (i.e. the second term) leads to a logarithmic confining potential
\cite{P2,robpoland}.

Similarly, one can show that the kinetic energy in the nonrelativistic
limit reduces to
\begin{eqnarray}
{\kappa^{\perp 2}+m_a^2\over{x_a}}+{\kappa^{\perp 2}+m_b^2\over{x_b}}
 & \rightarrow & 2 M_{ab} {\vec{k}^2\over{2 m}} \  \  ,
\end{eqnarray}
where $m$ is the reduced mass;
and that the self-energy produces only a constant shift:
\begin{eqnarray}
%\lefteqn{
\Sigma_a+\Sigma_b \rightarrow %}\nonumber\\& &
 &  &{\alpha C_F M_{ab} {\cal L}\over{ \pi}}
  \left[ 2 \log \left({P^+\over{\epsilon {\cal P}^+}}\right)
 +2 \log \left( {{\cal L}\over{M_{ab}}}\right) +
+{1\over{4}}{m_a\over{ {\cal L}+m_a}}+{1\over{4}}{m_b\over{ {\cal L}+m_b}}
\right. %}
\nonumber\\
  &  &
\left.
\left( 1+{3m_a\over{4{\cal L}}}\right)
 \log \left( {m_a\over{ {\cal L}+m_a}} \right)+
\left( 1+{3m_b\over{4{\cal L}}}\right)
\log \left( {m_b\over{ {\cal L}+m_b}}\right)
-{3\over{2}} \right] \  \  ,
\end{eqnarray}
where
 $${\cal L} \equiv  {\Lambda ^2 \over{{\cal P}^+}}{P^+\over{M_{ab}}} $$
carries dimension of mass and in the nonrelativistic limit replaces the
light-front cutoff ${\Lambda ^2 \over{{\cal P}^+}}$.
Note that
we did not assume anything about the relation between masses and the cutoff
${\cal  L}$.
The assumption that momenta are small in comparison to masses is equivalent
to assuming that the wave function is peaked at small momenta, which typically
requires that $g_{\Lambda} $ is sufficiently small.

\begin{figure}[th!]
\begin{center}
\input{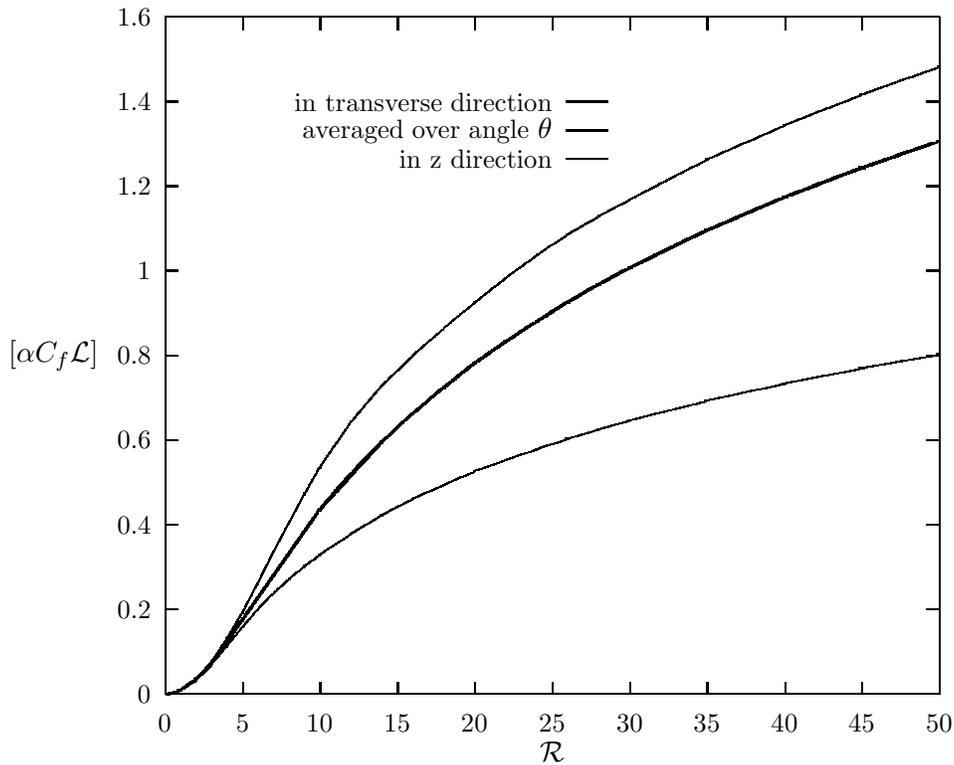}
\end{center}
%Figure 2.
\caption{Violation of rotational symmetry in the confining potential. At
small distances, the violation of rotational symmetry is small.  At
any
fixed value of ${\cal R}$, the
confining potential is maximal when the quarks are separated in purely
transverse direction. It is minimal when the separation between the quarks
is in purely longitudinal direction. We also show the strength of the
potential averaged over the angle $\theta$.}
\end{figure}
It is more intuitive to work in position space.
The Fourier transform of the potential~is:
\begin{eqnarray}
\lefteqn{2 M_{ab} \  V(\vec{r}) = -{2 M_{ab} C_F \alpha \over{r}} }\nonumber\\
 &  & - {2 M_{ab} C_F \alpha {\cal L}\over{\pi}} \int_0^1 dt \ {1-t\over{t}}
\left\{ cos(t{\cal L}r_z)\left[J_0(\sqrt{t-t^2}{\cal L}r_{\perp})+
J_2(\sqrt{t-t^2}{\cal L}r_{\perp})\right] -1\right\}\nonumber\\
 &  & +{2 M_{ab} C_F \alpha {\cal L}\over{\pi}} \int_0^{\infty}
{dt \ dw \over{t^2 +w}} \theta(t-t^2-w)
\left\{ cos(t{\cal L}r_z)J_0(\sqrt{w}{\cal L}r_{\perp})-1 \right\}
\end{eqnarray}
where $r_z$ is the z-component of the separation between the quarks and
$r_{\perp} = \vert \vec{r}_{\perp} \vert$ is the transverse separation
between the quarks.
 Integration variables $t$ and $w$ are dimensionless.
The first term in the expression is the Coulomb potential, the rest is
the confining potential normalized to zero at the origin. As mentioned above,
for the color singlet states, the infrared divergence in this two-body operator
precisely cancels with the infrared divergence in the
self-energies, making the confining potential finite at the origin.
Any finite terms required
to make the confining potential vanish at the origin are subtracted from
the self-energies.

The confining potential is not rotationally invariant, because the cutoff
 violates rotational symmetry.
 Also, recall that rotations on the light-front are
not kinematic, so as long as the gluon emission and absorption is allowed,
complete rotational invariance requires states containing arbitrarily large
numbers of gluons. The renormalized Hamiltonian restores
rotational symmetry, but only to ${\cal O}(g^2)$, while in the bound
state calculation the confining interaction is treated to all orders.

The expansion of the confining
 potential in Legendre polynomials has only even terms:
\begin{eqnarray}
V(\vec{r}) =  {\alpha C_F {\cal L}\over{\pi}}\sum_k V_{2k}(r) P_{2k}(cos\theta)
\  \  ,
\end{eqnarray}
with
\begin{eqnarray}
V_0  & = &
2 \log {\cal R}-2 Ci({\cal R}) +4 {Si({\cal R})\over{{\cal R}}}
-2 {(1-\cos{\cal R})\over{{\cal R}^2}}+2{\sin{\cal R}\over{{\cal R}}}
-5 +2\gamma  \  \
, \\
%\end{eqnarray}
%
%\begin{eqnarray}
V_2 &  = & - {5\over{3}} +{5 Si({\cal R})\over{{\cal R}}}
-{10\over{{\cal R}^2}}+{5 Si({\cal R})\over{2{\cal R}^2}}
+{5 cos {\cal R}\over{{\cal R}^2}}\nonumber\\
& \   & - {15\over{{\cal R}^3}} +{5 cos {\cal R}\over{2{\cal R}^3}}
+{5 sin {\cal R}\over{{\cal R}^3}}
+{5 sin {\cal R}\over{2{\cal R}^4}}
+{20\over{{\cal R}^5}} -{20 cos {\cal R}\over{{\cal R}^5}} \   \  ,
\end{eqnarray}
and
\begin{eqnarray*}
V_{2k} = {(4k+1)\over{2}}\int_{-1}^{+1} dx
\left[{\alpha C_F {\cal L}\over{\pi}}\right]^{-1} V(\vec{r}) P_{2k}(x) \  \
 ,
\end{eqnarray*}
\noindent
where ${\cal R}= {\cal L} r$; $Ci(x)$, $Si(x)$ are cosine and sine
integrals, respectively; and $\gamma$ is the Euler constant.

When the separation between the quarks is purely in the z-direction, i.e.
$r_{\perp} =0$, the potential has its minimum value with respect to the angle
$\theta $.
Integrals in $(21)$ can be done analytically leading to:
\begin{eqnarray}
{2 M_{ab} C_F \alpha {\cal L}\over{\pi}} \left[ \log({\cal R})
-Ci({\cal R}) + {sin({\cal R})\over{{\cal R}}} +{Si({\cal R})\over{{\cal R}}}
- (2 - \gamma)\right].
\end{eqnarray}

The potential is maximal in the purely transverse direction (i.e. $r_z=0$).
For large separations, the difference between the potential in purely
transverse and purely longitudinal directions is a factor of two.
At very short distances, even the
confining potential is rotationally invariant (see figure 2).
This is of no help though, because at very short distances, the Coulomb
part of the potential dominates.

For the bound state calculation, we want to choose a rotationally invariant
confining interaction for $H_0$.
This is partly motivated by phenomenology and partly by
our desire to find a tool to qualitatively analyse the effective operators
in the Hamiltonian. Restoration of rotational symmetry in $H_0$
provides such a tool in the form of a simple scaling analysis.

It is not clear how we should choose the leading rotationally invariant
interaction at order $g^2$, because we do not
 know yet how  rotational symmetry is restored.

It is clear from  $(24)$ that it is convenient to use the first
term in the expansion in Legendre polynomials. It is the only choice that
does not lead to any corrections in the first order of  bound state
perturbation theory. The corrections which come from the rotationally
noninvariant terms enter at second order in bound state perturbation theory;
that is, at the same order as the emission and absorption processes.
We have to mention that these corrections are not suppressed by powers of
the coupling, and that the problem does not disappear in the low-lying bound
states as the coupling decreases.
It all returns to the issue of how rotational symmetry is recovered in
this approach - an issue
 that extends beyond our simple leading order calculation.
\vskip0.3in

We will conclude this section with a list of approximations we make in the
second step of the calculation:

\begin{enumerate}
\item{ We do not include emission and absorption in  $H_0$.}

\item{ We replace $x_a$, $x_b=1-x_a$ by $x_a= {m_a\over{M_{ab}}} - \eta$,
 $1-x_a= {m_b\over{M_{ab}}} + \eta$; and Taylor expand ${1\over{x}}$,
${1\over{1-x}}$ in powers of $\eta$.

In energy denominators, we neglect terms that are higher than second order
in momenta.
The same approximation is made in the arguments of step functions. This
leads
to a ``new '' cutoff ${\cal L}$, which carries only one power of transverse
mass. It
should be emphasized that the new cutoff arises only in the second step, that
is, in the nonrelativistic approximation to the light-front effective
Hamiltonian. Only in this context does it replace the original cutoff
${\Lambda^2 \over{{\cal P}^+}}$. Elsewhere we have to work with
${\Lambda^2 \over{{\cal P}^+}}$ (e.g. the coupling runs with
${\Lambda^2 \over{{\cal P}^+}}$). It is $\Lambda^2$ which has to be at hadronic
mass scales.
In the self-energies we keep only the leading constant shift,
 which is independent
of momenta. This is because the self-energies are already ${\cal O}(g^2)$.}

\item{ We introduce the third component of the ``equal-time'' momentum:
$k_z = \pm M_{ab} \eta$, and extend the integration limits on $q_z$ from
$(-M_{ab}, +M_{ab})$ to $(-\infty, +\infty)$. It is easy to show
that this third
component of the equal-time momenta coincides with the
z-component of equal-time momentum in the lowest order in $\alpha$.
As mentioned earlier, it is
possible to introduce a third component without nonrelativistic reduction
\cite{coester}, but it does not yield any simplifications if the masses
are not
equal. In the nonrelativistic limit they both agree to the leading order
in powers of momenta.
The main point we want to make here is that in the nonrelativistic limit
the dynamics naturally lead to equal-time dynamics.}

\item{ For the leading order bound state calculation, we include only the
rotationally invariant moment of the confining potential (i.e. $V_0$ given in
$(25)$ ) in $H_0$.}
\end{enumerate}

\vskip0.3in

This gives $H_0$ which we choose
for the purposes of the bound state calculation:
\begin{eqnarray}
H_0 = 2M_{ab}\left[ -{1\over{2m}}\vec{\nabla}^2 +\tilde{\Sigma}
- { C_F \alpha \over{r}} +{ C_F \alpha {\cal L} \over{\pi}} V_0({\cal L}r)
\right],
\end{eqnarray}
\noindent where $V_0({\cal L}r)$ is given in $(25)$ and
 $\tilde{\Sigma }$ contains the finite shift produced by self-energies
after subtracting terms needed to make the confining potential vanish at
the origin:
\begin{eqnarray}
\tilde{\Sigma } &  =  &
 {\alpha C_F  {\cal L}\over{ 2 \pi}}
\left[
\left( 1+{3m_a\over{4{\cal L}}}\right)
 \log \left( {m_a\over{ {\cal L}+m_a}} \right)  +
\left( 1+{3m_b\over{4{\cal L}}}\right)
\log \left( {m_b\over{ {\cal L}+m_b}}\right) \right.% }
\nonumber\\
  & \ \ \   &  \  \  \  \  \  \  \  \  \  \   \left.
+{1\over{4}}{m_a\over{ {\cal L}+m_a}}+{1\over{4}}{m_b\over{ {\cal L}+m_b}}
+{5\over{2}}
\right]  .
\end{eqnarray}
%\vfill
%\pagebreak

\subsection{Schr{\H{o}}dinger equation.}

We now want to find the mass of a bound state and its wave function
$\psi (\kappa^{\perp}, x)$:
\begin{eqnarray}
\vert P \rangle = \int {d^2\kappa^{\perp} \ dx \over{2(2\pi)^3
\sqrt{x(1-x)}}}\psi(\kappa^{\perp},x)
b^{\dagger} d^{\dagger} \vert 0 \rangle \  \  .
\end{eqnarray}
We use  Lorentz  invariant normalization for the states:
$$\langle P' \vert P \rangle = 2(2\pi)^3 P^+ \delta ^3(\vec{P}-\vec{P'}),$$
and the wave function is normalized to one:
$$  \int {d^2\kappa^{\perp} \ dx \over{2(2\pi)^3
}}\vert \psi(\kappa^{\perp},x) \vert^2 = 1 .$$
The bound state satisfies:
\begin{eqnarray}
H_{0} \vert P \rangle = {\cal M}^2 \vert P \rangle \  \  ,
\end{eqnarray}
where ${\cal M}^2$ is the invariant mass of the bound state.
Let the mass of the bound state be
\begin{eqnarray}
{\cal M}^2 = (m_a +m_b)^2 +2(m_a+m_b) E \   \  ,
\end{eqnarray}
which defines $E$.

Substituting for  $H_0$ and ${\cal M}^2$ in
equation $(31)$, after some straightforward algebra one obtains a bound
state equation for the wave function $\psi$:
\begin{eqnarray}
 M_{ab} \left(E - \tilde{\Sigma}
 + {1\over{2m}}{d^2\over{d\vec{r}^2}}\right) \psi (\vec{r}) =
 M_{ab} \left[{-\alpha C_F \over{r}}+{\alpha C_F {\cal L}\over{\pi}}
V_{\rm conf}({\cal L}\vec{r})
\right]\psi (\vec{r}) \  \  ,
\end{eqnarray}
where $m$ is the reduced mass and we choose
$V_{\rm conf}({\cal L}\vec{r})=V_0({\cal L}r)$ given in $(25)$.
We choose the confining potential for the leading order
bound state calculation to be rotationally invariant, but
other choices are possible. To remind the reader of this possibility and
to keep the discussion as general as possible, we use $\vec{{\cal R}}$
as the argument of the confining potential.

It is convenient to use a dimensionless separation ${\cal R}={\cal L}r$ that
naturally arises in the confining piece of the potential, and to absorb
$-\tilde{\Sigma}$
into a definition of the
eigenvalue $\tilde{E}$ of the Schr{\H{o}}dinger equation. When extracting the
bound state mass, $-\tilde{\Sigma}$  has to be subtracted.
The bound state equation in the  dimensionless separation is:
\begin{eqnarray}
\left[ -{{\cal L}^2\over{2m}}{d^2\over{d\vec{{\cal R}}^2}}
+ {\cal L} \alpha C_F \left( {1\over{\pi}}V_{\rm conf}(\vec{\cal R}) + V_{\rm
coul}(
{\cal R})\right) \right] \psi({\vec{{\cal R}}}) =
 \tilde{E}\psi({\vec{{\cal R}}}) \  \  .
\end{eqnarray}
Multiplying both sides of the equation by $2m/{\cal L}^2$
and introducing a dimensionless coupling and eigenvalue:
\begin{eqnarray}
c & \equiv & {2m\alpha C_F \over{{\cal L}}} ,
\\
e & \equiv & {2m \tilde{E} \over{{\cal L}^2}} ,
\end{eqnarray}
we obtain a Schr{\H{o}}dinger equation, which depends only on dimensionless
variables:
\begin{eqnarray}
\left[ -{d^2\over{d\vec{{\cal R}}^2}}
+ c \left(
 {1\over{\pi}}V_{\rm conf}(\vec{\cal R}) + V_{\rm coul}(
{\cal R})\right) \right] \psi({\vec{{\cal R}}}) =
 e \psi({\vec{{\cal R}}}) \  \  .
\end{eqnarray}

This form is advantageous for numerical study, but moreover, it is
quite general
- one obtains an equation of this form for any quark-antiquark
systems and any choice of the confining potential
in the nonrelativistic limit, regardless of the masses, providing
they are nonzero.
For different systems ${\cal L}$, $c$, $e$ would differ,
 but the resulting dimensionless
Schr{\H{o}}dinger
equation will be the same. Thus in the leading order,  {\it qualitative}
characteristics of spectra depend only on one particular combination of the
masses
and the coupling.
\vfill
\pagebreak
\subsection{Bohr analysis.}

The purpose of this analysis is to gain a qualitative understanding of the
physics described by the Schr{\H{o}}dinger eqn. $(37)$.

Since this is only  a qualitative analysis, we will neglect the finite terms
in the confining potential, but keep in mind that at small distances the
confining potential vanishes. Since the Schr{\H{o}}dinger equation is
dimensionless, all quantities in this analysis are dimensionless also.
The eigenvalue (energy $e$) is given by a
sum of the kinetic
energy, which in our case is simply $p^2$, $p$ being dimensionless; and
the potential energy. We use the
uncertainty principle to replace the  momentum by
${1\over{{\cal R}}}$.

Let us consider the ground state:
\begin{eqnarray}
e_0 \approx p^2 + V({\cal R}) \approx {1\over{{\cal R}^2}} + c (2 \log{\cal R}
- {1\over{{\cal R}}}) \  \  .
\end{eqnarray}
Now we  find ${\cal R}$ which minimizes the energy:
\begin{eqnarray*}
{de_0 \over{d{\cal R}}} =0 \  \  .
\end{eqnarray*}
The solution is
\begin{eqnarray*}
{\cal R}_0 = {\sqrt{c^2 +8c} -c\over{2c}} \  \   .
\end{eqnarray*}
Similarly, we find the $l=1$ excited state, for which
\begin{eqnarray*}
{\cal R}_1 =  {\sqrt{c^2 +3\cdot 8c} -c\over{2c}} \  \  .
\end{eqnarray*}

We now consider two limiting cases: when $c$ is small and when $c$ is large.

{\bf c small:} In this case, in the ground state
${\cal R}_0 \approx {1\over{\sqrt{c}}}$,  and the energy
 $e_0 \approx c+ c\  \log {1\over{\sqrt{c}}}$. In the lowest $l=1$ state
${{\cal R}_1 /{{\cal R_0}}} \approx \sqrt{3}$,
and the splitting in the energy
between these two states is $e_1 -e_0 \approx c \ \log \sqrt{3}$.

Of course, we are interested in ``real world'' energies and not the
dimensionless results. If we ``unwrap'' the dimensionless results,
\begin{eqnarray}
e_1 - e_0 &  \approx &{m\over{ {\cal L}^2}} (E_1 -E_0) \nonumber\\
\approx c &  \approx & {m\over{ {\cal L}}} \alpha
\end{eqnarray}
which implies that
\begin{eqnarray}
E_1 -E_0 \approx \alpha {\cal L} .
\end{eqnarray}
The splitting between the ground state and the lowest lying P-state is
independent of masses.

Similarly one can show that what sets the scale for individual energies is
${\cal L}$ and that the size of the system, ${\cal R}$, depends both on
${\cal L}$ and the reduced mass $m$.

{\bf c large}: In this case, ${\cal R}_0$ scales like $2\over{c}$,
the ground state energy $e_0 \approx -{c^2\over{4}}$, while
${\cal R}_1 \approx {6\over{c}}$ and
$e_1 \approx -{c^2\over{4}}{1\over{6}}$.

Unwrapping shows that the scale ${\cal L}$ drops out:
\begin{eqnarray}
e & \approx & {m \over{{\cal L}^2}}E \nonumber\\
\approx c^2 & \approx & ({m \over{{\cal L}}}\alpha )^2 ,
\end{eqnarray}
where $e$ stands for any of the dimensionless energies under
consideration, and $E$ for any of the ``real'' bindings:
\begin{eqnarray}
E \approx \alpha^2 m
\end{eqnarray}
which depend only on the reduced mass but not on the confining scale.

Similarly, ${\cal L}$ drops out also of the expression for the size.

In summary, when $c$ is small, the spectra are determined by the logarithmic
part of the potential; when $c$ is large, the spectra are Coulombic.
Recall that $c$ is proportional to the reduced mass.
Let us set aside questions on how ${\cal L}$, or the original light-front
$\Lambda^2$, depends on the masses of constituents.
There is a natural
distinction between heavy quarkonia and other mesonic systems. In the case of
heavy quarkonia the reduced mass is always proportional to
the heavy mass, while in all other
systems it is related to the light mass. This means that $c$ is  larger
for the heavy quarkonia, and thus these states are substantially more
 Coulombic; while the other systems, including heavy mesons, are more
influenced by the confining potential.

This simple technique can be used to estimate expectation values of various
operators in the ground state.
%\vfill
%\pagebreak

\begin{figure}[th!]
\begin{center}
% GNUPLOT: LaTeX picture
\setlength{\unitlength}{0.240900pt}
\ifx\plotpoint\undefined\newsavebox{\plotpoint}\fi
\sbox{\plotpoint}{\rule[-0.200pt]{0.400pt}{0.400pt}}%
\begin{picture}(1500,1200)(0,0)
\font\gnuplot=cmr10 at 10pt
\gnuplot
\sbox{\plotpoint}{\rule[-0.200pt]{0.400pt}{0.400pt}}%
\put(220.0,566.0){\rule[-0.200pt]{292.934pt}{0.400pt}}
\put(220.0,113.0){\rule[-0.200pt]{0.400pt}{245.477pt}}
\put(220.0,113.0){\rule[-0.200pt]{4.818pt}{0.400pt}}
\put(198,113){\makebox(0,0)[r]{-0.2}}
\put(1416.0,113.0){\rule[-0.200pt]{4.818pt}{0.400pt}}
\put(220.0,226.0){\rule[-0.200pt]{4.818pt}{0.400pt}}
\put(198,226){\makebox(0,0)[r]{-0.15}}
\put(1416.0,226.0){\rule[-0.200pt]{4.818pt}{0.400pt}}
\put(220.0,339.0){\rule[-0.200pt]{4.818pt}{0.400pt}}
\put(198,339){\makebox(0,0)[r]{-0.1}}
\put(1416.0,339.0){\rule[-0.200pt]{4.818pt}{0.400pt}}
\put(220.0,453.0){\rule[-0.200pt]{4.818pt}{0.400pt}}
\put(198,453){\makebox(0,0)[r]{-0.05}}
\put(1416.0,453.0){\rule[-0.200pt]{4.818pt}{0.400pt}}
\put(220.0,566.0){\rule[-0.200pt]{4.818pt}{0.400pt}}
\put(198,566){\makebox(0,0)[r]{0}}
\put(1416.0,566.0){\rule[-0.200pt]{4.818pt}{0.400pt}}
\put(220.0,679.0){\rule[-0.200pt]{4.818pt}{0.400pt}}
\put(198,679){\makebox(0,0)[r]{0.05}}
\put(1416.0,679.0){\rule[-0.200pt]{4.818pt}{0.400pt}}
\put(220.0,792.0){\rule[-0.200pt]{4.818pt}{0.400pt}}
\put(198,792){\makebox(0,0)[r]{0.1}}
\put(1416.0,792.0){\rule[-0.200pt]{4.818pt}{0.400pt}}
\put(220.0,906.0){\rule[-0.200pt]{4.818pt}{0.400pt}}
\put(198,906){\makebox(0,0)[r]{0.15}}
\put(1416.0,906.0){\rule[-0.200pt]{4.818pt}{0.400pt}}
\put(220.0,1019.0){\rule[-0.200pt]{4.818pt}{0.400pt}}
\put(198,1019){\makebox(0,0)[r]{0.2}}
\put(1416.0,1019.0){\rule[-0.200pt]{4.818pt}{0.400pt}}
\put(220.0,1132.0){\rule[-0.200pt]{4.818pt}{0.400pt}}
\put(198,1132){\makebox(0,0)[r]{0.25}}
\put(1416.0,1132.0){\rule[-0.200pt]{4.818pt}{0.400pt}}
\put(220.0,113.0){\rule[-0.200pt]{0.400pt}{4.818pt}}
\put(220,68){\makebox(0,0){0}}
\put(220.0,1112.0){\rule[-0.200pt]{0.400pt}{4.818pt}}
\put(463.0,113.0){\rule[-0.200pt]{0.400pt}{4.818pt}}
\put(463,68){\makebox(0,0){0.2}}
\put(463.0,1112.0){\rule[-0.200pt]{0.400pt}{4.818pt}}
\put(706.0,113.0){\rule[-0.200pt]{0.400pt}{4.818pt}}
\put(706,68){\makebox(0,0){0.4}}
\put(706.0,1112.0){\rule[-0.200pt]{0.400pt}{4.818pt}}
\put(950.0,113.0){\rule[-0.200pt]{0.400pt}{4.818pt}}
\put(950,68){\makebox(0,0){0.6}}
\put(950.0,1112.0){\rule[-0.200pt]{0.400pt}{4.818pt}}
\put(1193.0,113.0){\rule[-0.200pt]{0.400pt}{4.818pt}}
\put(1193,68){\makebox(0,0){0.8}}
\put(1193.0,1112.0){\rule[-0.200pt]{0.400pt}{4.818pt}}
\put(1436.0,113.0){\rule[-0.200pt]{0.400pt}{4.818pt}}
\put(1436,68){\makebox(0,0){1}}
\put(1436.0,1112.0){\rule[-0.200pt]{0.400pt}{4.818pt}}
\put(220.0,113.0){\rule[-0.200pt]{292.934pt}{0.400pt}}
\put(1436.0,113.0){\rule[-0.200pt]{0.400pt}{245.477pt}}
\put(220.0,1132.0){\rule[-0.200pt]{292.934pt}{0.400pt}}
\put(45,622){\makebox(0,0){$e$}}
\put(828,23){\makebox(0,0){$c$}}
\put(828,1177){\makebox(0,0){ }}
\put(220.0,113.0){\rule[-0.200pt]{0.400pt}{245.477pt}}
\sbox{\plotpoint}{\rule[-0.500pt]{1.000pt}{1.000pt}}%
\put(1193,792){\makebox(0,0)[r]{l=0}}
\multiput(1215,792)(20.756,0.000){4}{\usebox{\plotpoint}}
\put(1281,792){\usebox{\plotpoint}}
\put(256,651){\usebox{\plotpoint}}
\multiput(256,651)(10.925,17.648){2}{\usebox{\plotpoint}}
\put(277.98,686.22){\usebox{\plotpoint}}
\put(289.37,703.56){\usebox{\plotpoint}}
\multiput(293,709)(12.453,16.604){2}{\usebox{\plotpoint}}
\multiput(317,741)(13.552,15.720){2}{\usebox{\plotpoint}}
\multiput(342,770)(14.314,15.030){4}{\usebox{\plotpoint}}
\multiput(402,833)(15.795,13.465){4}{\usebox{\plotpoint}}
\multiput(463,885)(16.833,12.142){3}{\usebox{\plotpoint}}
\multiput(524,929)(17.746,10.764){4}{\usebox{\plotpoint}}
\multiput(585,966)(18.595,9.221){6}{\usebox{\plotpoint}}
\multiput(706,1026)(19.575,6.899){6}{\usebox{\plotpoint}}
\multiput(828,1069)(20.116,5.112){7}{\usebox{\plotpoint}}
\multiput(950,1100)(20.504,3.220){5}{\usebox{\plotpoint}}
\multiput(1071,1119)(20.699,1.527){6}{\usebox{\plotpoint}}
\multiput(1193,1128)(20.755,0.172){6}{\usebox{\plotpoint}}
\multiput(1314,1129)(20.721,-1.189){6}{\usebox{\plotpoint}}
\put(1436,1122){\usebox{\plotpoint}}
\sbox{\plotpoint}{\rule[-0.200pt]{0.400pt}{0.400pt}}%
\put(1193,747){\makebox(0,0)[r]{l=1}}
\put(1215.0,747.0){\rule[-0.200pt]{15.899pt}{0.400pt}}
\put(256,642){\usebox{\plotpoint}}
\multiput(256.58,642.00)(0.493,0.695){23}{\rule{0.119pt}{0.654pt}}
\multiput(255.17,642.00)(13.000,16.643){2}{\rule{0.400pt}{0.327pt}}
\multiput(269.58,660.00)(0.492,0.669){21}{\rule{0.119pt}{0.633pt}}
\multiput(268.17,660.00)(12.000,14.685){2}{\rule{0.400pt}{0.317pt}}
\multiput(281.58,676.00)(0.492,0.625){21}{\rule{0.119pt}{0.600pt}}
\multiput(280.17,676.00)(12.000,13.755){2}{\rule{0.400pt}{0.300pt}}
\multiput(293.58,691.00)(0.496,0.562){45}{\rule{0.120pt}{0.550pt}}
\multiput(292.17,691.00)(24.000,25.858){2}{\rule{0.400pt}{0.275pt}}
\multiput(317.00,718.58)(0.519,0.496){45}{\rule{0.517pt}{0.120pt}}
\multiput(317.00,717.17)(23.928,24.000){2}{\rule{0.258pt}{0.400pt}}
\multiput(342.00,742.58)(0.612,0.498){95}{\rule{0.590pt}{0.120pt}}
\multiput(342.00,741.17)(58.776,49.000){2}{\rule{0.295pt}{0.400pt}}
\multiput(402.00,791.58)(0.764,0.498){77}{\rule{0.710pt}{0.120pt}}
\multiput(402.00,790.17)(59.526,40.000){2}{\rule{0.355pt}{0.400pt}}
\multiput(463.00,831.58)(0.957,0.497){61}{\rule{0.863pt}{0.120pt}}
\multiput(463.00,830.17)(59.210,32.000){2}{\rule{0.431pt}{0.400pt}}
\multiput(524.00,863.58)(1.137,0.497){51}{\rule{1.004pt}{0.120pt}}
\multiput(524.00,862.17)(58.917,27.000){2}{\rule{0.502pt}{0.400pt}}
\multiput(585.00,890.58)(1.560,0.498){75}{\rule{1.341pt}{0.120pt}}
\multiput(585.00,889.17)(118.217,39.000){2}{\rule{0.671pt}{0.400pt}}
\multiput(706.00,929.58)(2.371,0.497){49}{\rule{1.977pt}{0.120pt}}
\multiput(706.00,928.17)(117.897,26.000){2}{\rule{0.988pt}{0.400pt}}
\multiput(828.00,955.58)(4.157,0.494){27}{\rule{3.353pt}{0.119pt}}
\multiput(828.00,954.17)(115.040,15.000){2}{\rule{1.677pt}{0.400pt}}
\multiput(950.00,970.59)(10.887,0.482){9}{\rule{8.167pt}{0.116pt}}
\multiput(950.00,969.17)(104.050,6.000){2}{\rule{4.083pt}{0.400pt}}
\put(1071,974.17){\rule{24.500pt}{0.400pt}}
\multiput(1071.00,975.17)(71.149,-2.000){2}{\rule{12.250pt}{0.400pt}}
\multiput(1193.00,972.93)(7.956,-0.488){13}{\rule{6.150pt}{0.117pt}}
\multiput(1193.00,973.17)(108.235,-8.000){2}{\rule{3.075pt}{0.400pt}}
\multiput(1314.00,964.92)(4.157,-0.494){27}{\rule{3.353pt}{0.119pt}}
\multiput(1314.00,965.17)(115.040,-15.000){2}{\rule{1.677pt}{0.400pt}}
\sbox{\plotpoint}{\rule[-0.400pt]{0.800pt}{0.800pt}}%
\put(1193,702){\makebox(0,0)[r]{ground state}}
\put(1215.0,702.0){\rule[-0.400pt]{15.899pt}{0.800pt}}
\put(256,619){\usebox{\plotpoint}}
\multiput(256.00,620.40)(0.589,0.512){15}{\rule{1.145pt}{0.123pt}}
\multiput(256.00,617.34)(10.623,11.000){2}{\rule{0.573pt}{0.800pt}}
\multiput(269.00,631.40)(0.674,0.516){11}{\rule{1.267pt}{0.124pt}}
\multiput(269.00,628.34)(9.371,9.000){2}{\rule{0.633pt}{0.800pt}}
\multiput(281.00,640.40)(0.913,0.526){7}{\rule{1.571pt}{0.127pt}}
\multiput(281.00,637.34)(8.738,7.000){2}{\rule{0.786pt}{0.800pt}}
\multiput(293.00,647.41)(0.947,0.509){19}{\rule{1.677pt}{0.123pt}}
\multiput(293.00,644.34)(20.519,13.000){2}{\rule{0.838pt}{0.800pt}}
\multiput(317.00,660.40)(1.322,0.514){13}{\rule{2.200pt}{0.124pt}}
\multiput(317.00,657.34)(20.434,10.000){2}{\rule{1.100pt}{0.800pt}}
\multiput(342.00,670.41)(2.086,0.508){23}{\rule{3.400pt}{0.122pt}}
\multiput(342.00,667.34)(52.943,15.000){2}{\rule{1.700pt}{0.800pt}}
\multiput(402.00,685.38)(9.828,0.560){3}{\rule{9.960pt}{0.135pt}}
\multiput(402.00,682.34)(40.328,5.000){2}{\rule{4.980pt}{0.800pt}}
\put(463,686.34){\rule{14.695pt}{0.800pt}}
\multiput(463.00,687.34)(30.500,-2.000){2}{\rule{7.347pt}{0.800pt}}
\multiput(524.00,685.08)(3.766,-0.516){11}{\rule{5.622pt}{0.124pt}}
\multiput(524.00,685.34)(49.331,-9.000){2}{\rule{2.811pt}{0.800pt}}
\multiput(585.00,676.09)(1.985,-0.503){55}{\rule{3.323pt}{0.121pt}}
\multiput(585.00,676.34)(114.104,-31.000){2}{\rule{1.661pt}{0.800pt}}
\multiput(706.00,645.09)(1.281,-0.502){89}{\rule{2.233pt}{0.121pt}}
\multiput(706.00,645.34)(117.365,-48.000){2}{\rule{1.117pt}{0.800pt}}
\multiput(828.00,597.09)(0.957,-0.501){121}{\rule{1.725pt}{0.121pt}}
\multiput(828.00,597.34)(118.420,-64.000){2}{\rule{0.863pt}{0.800pt}}
\multiput(950.00,533.09)(0.798,-0.501){145}{\rule{1.474pt}{0.121pt}}
\multiput(950.00,533.34)(117.941,-76.000){2}{\rule{0.737pt}{0.800pt}}
\multiput(1071.00,457.09)(0.686,-0.501){171}{\rule{1.297pt}{0.121pt}}
\multiput(1071.00,457.34)(119.309,-89.000){2}{\rule{0.648pt}{0.800pt}}
\multiput(1193.00,368.09)(0.593,-0.501){197}{\rule{1.149pt}{0.121pt}}
\multiput(1193.00,368.34)(118.615,-102.000){2}{\rule{0.575pt}{0.800pt}}
\multiput(1314.00,266.09)(0.539,-0.501){219}{\rule{1.064pt}{0.121pt}}
\multiput(1314.00,266.34)(119.792,-113.000){2}{\rule{0.532pt}{0.800pt}}
\end{picture}
\end{center}
%Figure 3.
\caption{Dimensionless eigenvalue $e$ for the ground state, the lowest
lying P state (l=1) and the lowest lying excited S state (l=0).}
\end{figure}
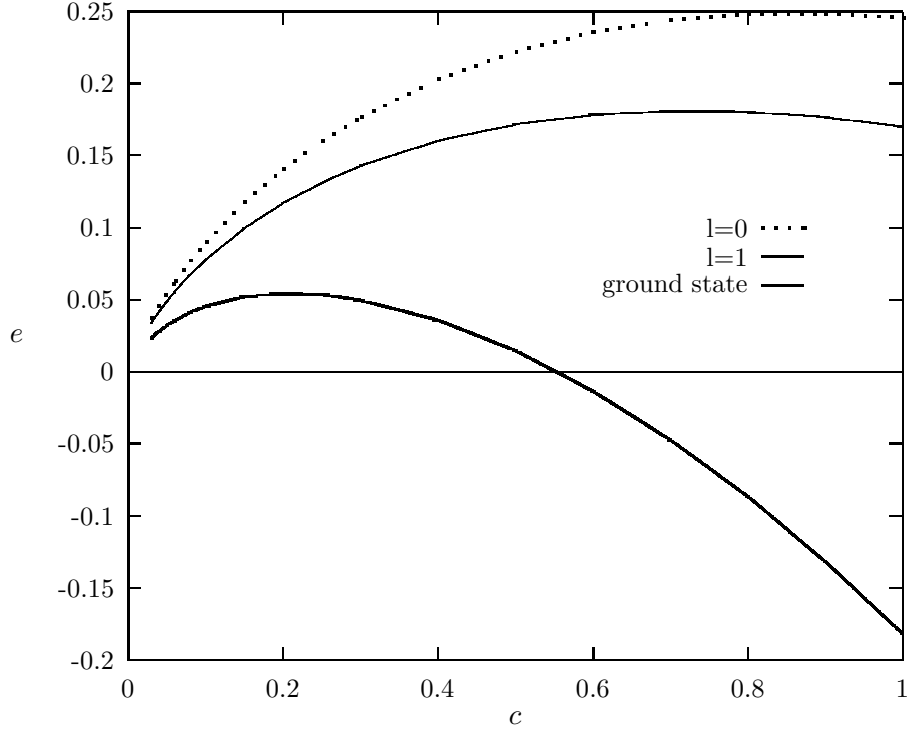

\subsection{Numerical results}

The numerical results are in agreement with the simple Bohr analysis.
Figure 3
presents the numerical results for the dimensionless eigenvalue.
It confirms that when $c$ is small, the spectra are dominated by the log
potential. As $c$ increases, the  Coulomb potential
becomes more important, especially for the
ground state.

\begin{figure}[th!]
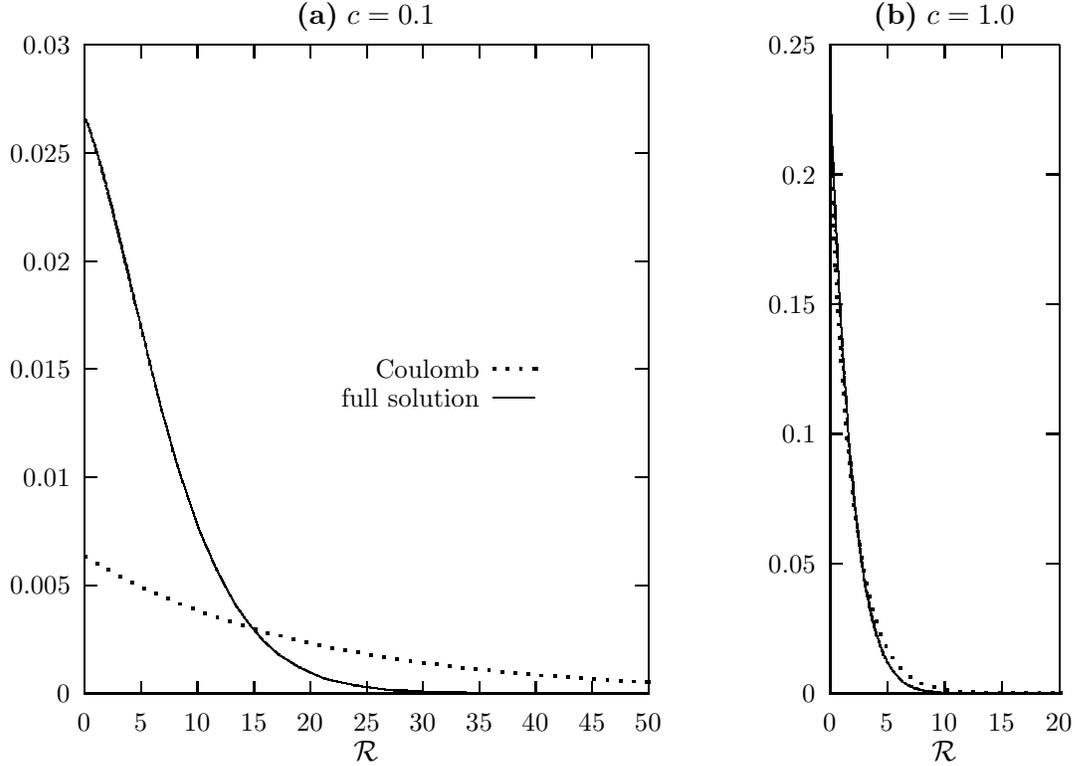

\input{fi4} \input{fi4b}
\caption{Ground state wavefunction compared to the Coulomb ground state
wavefunction at the same coupling. (a) For c=0.1, the ground state
wavefunction differs significantly from the Coulombic ground state
wavefunction, while (b) for c=1.0 they are similar.}
\end{figure}

Note that in the logarithmic regime (i.e., when $c$ is small) the ground state
dimensionless energy is always larger or at least comparable to the splitting
between the ground
  state and the $l=1$ state. For heavy mesons the
splitting is a few hundred MeV, which would imply that the binding in the
ground state is also large. However, recall that we had to absorb a
constant shift from the self-energies into the definition of the eigenvalue
%\noindent
and that the self-energy may be fine tuned  at low energies.
\setcounter{figure}{3}

\begin{figure}[th!]
\begin{center}
\input{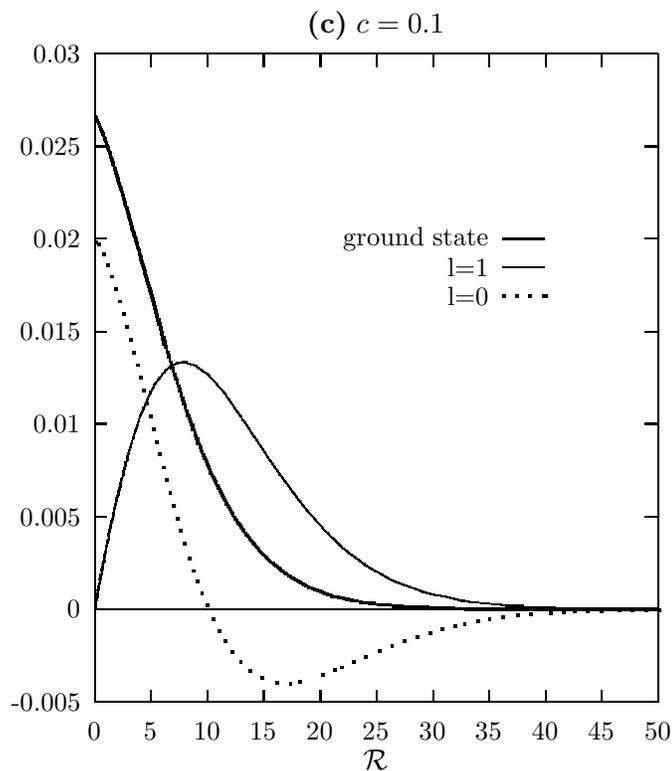}
\end{center}
\caption{(c) Wavefunctions of the ground state, the lowest
lying P state (l=1) and the lowest lying excited S state (l=0).}
\end{figure}

Figure 4
shows a few typical wavefunctions for the ground state
and lowest lying excited states.  In figure 4(a), (b) we compare
the exact solution of the Schr{\H{o}}dinger equation $(37)$ to the ground
state wavefunction of the Coulomb problem at the same value
of the coupling $c$.
When $c$ is small, the wavefunction is quite
different from the Coulombic wavefunction. As $c$ increases, it becomes closer
 and closer to Coulombic, in agreement with the  Bohr analysis presented above.
Wavefunctions of the lowest two excited states for $c=0.1$ which we later use
 to fit data, are shown in figure 4(c) together with the ground state
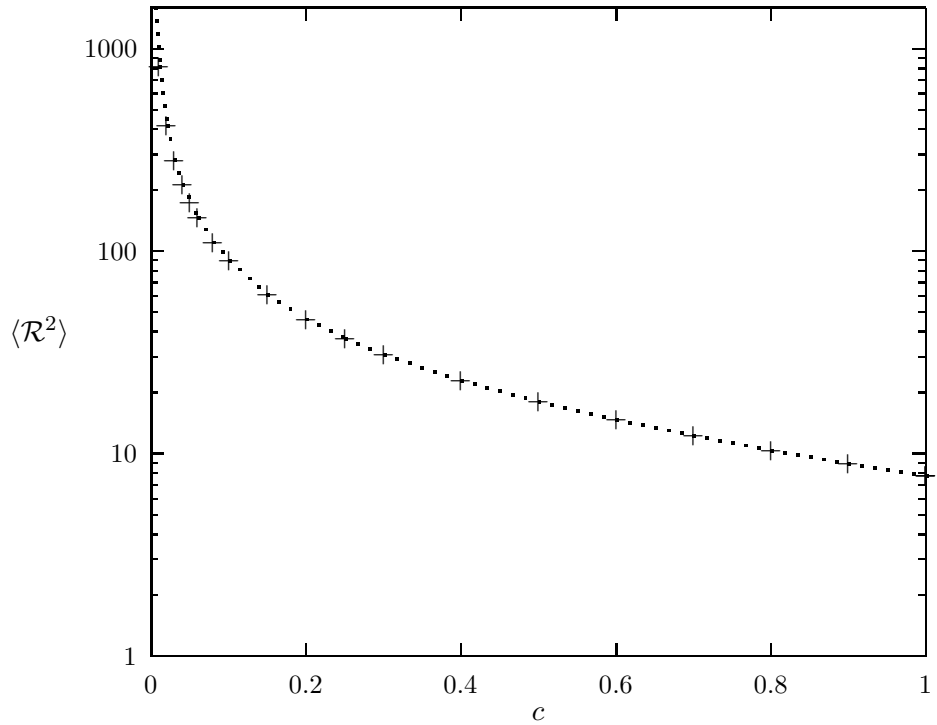
\begin{figure}[th!]
\begin{center}
% GNUPLOT: LaTeX picture
\setlength{\unitlength}{0.240900pt}
\ifx\plotpoint\undefined\newsavebox{\plotpoint}\fi
\sbox{\plotpoint}{\rule[-0.200pt]{0.400pt}{0.400pt}}%
\begin{picture}(1500,1200)(0,0)
\font\gnuplot=cmr10 at 10pt
\gnuplot
\sbox{\plotpoint}{\rule[-0.200pt]{0.400pt}{0.400pt}}%
\put(220.0,113.0){\rule[-0.200pt]{0.400pt}{245.477pt}}
\put(220.0,113.0){\rule[-0.200pt]{4.818pt}{0.400pt}}
\put(198,113){\makebox(0,0)[r]{1}}
\put(1416.0,113.0){\rule[-0.200pt]{4.818pt}{0.400pt}}
\put(220.0,209.0){\rule[-0.200pt]{2.409pt}{0.400pt}}
\put(1426.0,209.0){\rule[-0.200pt]{2.409pt}{0.400pt}}
\put(220.0,265.0){\rule[-0.200pt]{2.409pt}{0.400pt}}
\put(1426.0,265.0){\rule[-0.200pt]{2.409pt}{0.400pt}}
\put(220.0,304.0){\rule[-0.200pt]{2.409pt}{0.400pt}}
\put(1426.0,304.0){\rule[-0.200pt]{2.409pt}{0.400pt}}
\put(220.0,335.0){\rule[-0.200pt]{2.409pt}{0.400pt}}
\put(1426.0,335.0){\rule[-0.200pt]{2.409pt}{0.400pt}}
\put(220.0,360.0){\rule[-0.200pt]{2.409pt}{0.400pt}}
\put(1426.0,360.0){\rule[-0.200pt]{2.409pt}{0.400pt}}
\put(220.0,382.0){\rule[-0.200pt]{2.409pt}{0.400pt}}
\put(1426.0,382.0){\rule[-0.200pt]{2.409pt}{0.400pt}}
\put(220.0,400.0){\rule[-0.200pt]{2.409pt}{0.400pt}}
\put(1426.0,400.0){\rule[-0.200pt]{2.409pt}{0.400pt}}
\put(220.0,416.0){\rule[-0.200pt]{2.409pt}{0.400pt}}
\put(1426.0,416.0){\rule[-0.200pt]{2.409pt}{0.400pt}}
\put(220.0,431.0){\rule[-0.200pt]{4.818pt}{0.400pt}}
\put(198,431){\makebox(0,0)[r]{10}}
\put(1416.0,431.0){\rule[-0.200pt]{4.818pt}{0.400pt}}
\put(220.0,527.0){\rule[-0.200pt]{2.409pt}{0.400pt}}
\put(1426.0,527.0){\rule[-0.200pt]{2.409pt}{0.400pt}}
\put(220.0,583.0){\rule[-0.200pt]{2.409pt}{0.400pt}}
\put(1426.0,583.0){\rule[-0.200pt]{2.409pt}{0.400pt}}
\put(220.0,623.0){\rule[-0.200pt]{2.409pt}{0.400pt}}
\put(1426.0,623.0){\rule[-0.200pt]{2.409pt}{0.400pt}}
\put(220.0,653.0){\rule[-0.200pt]{2.409pt}{0.400pt}}
\put(1426.0,653.0){\rule[-0.200pt]{2.409pt}{0.400pt}}
\put(220.0,679.0){\rule[-0.200pt]{2.409pt}{0.400pt}}
\put(1426.0,679.0){\rule[-0.200pt]{2.409pt}{0.400pt}}
\put(220.0,700.0){\rule[-0.200pt]{2.409pt}{0.400pt}}
\put(1426.0,700.0){\rule[-0.200pt]{2.409pt}{0.400pt}}
\put(220.0,718.0){\rule[-0.200pt]{2.409pt}{0.400pt}}
\put(1426.0,718.0){\rule[-0.200pt]{2.409pt}{0.400pt}}
\put(220.0,735.0){\rule[-0.200pt]{2.409pt}{0.400pt}}
\put(1426.0,735.0){\rule[-0.200pt]{2.409pt}{0.400pt}}
\put(220.0,749.0){\rule[-0.200pt]{4.818pt}{0.400pt}}
\put(198,749){\makebox(0,0)[r]{100}}
\put(1416.0,749.0){\rule[-0.200pt]{4.818pt}{0.400pt}}
\put(220.0,845.0){\rule[-0.200pt]{2.409pt}{0.400pt}}
\put(1426.0,845.0){\rule[-0.200pt]{2.409pt}{0.400pt}}
\put(220.0,901.0){\rule[-0.200pt]{2.409pt}{0.400pt}}
\put(1426.0,901.0){\rule[-0.200pt]{2.409pt}{0.400pt}}
\put(220.0,941.0){\rule[-0.200pt]{2.409pt}{0.400pt}}
\put(1426.0,941.0){\rule[-0.200pt]{2.409pt}{0.400pt}}
\put(220.0,971.0){\rule[-0.200pt]{2.409pt}{0.400pt}}
\put(1426.0,971.0){\rule[-0.200pt]{2.409pt}{0.400pt}}
\put(220.0,997.0){\rule[-0.200pt]{2.409pt}{0.400pt}}
\put(1426.0,997.0){\rule[-0.200pt]{2.409pt}{0.400pt}}
\put(220.0,1018.0){\rule[-0.200pt]{2.409pt}{0.400pt}}
\put(1426.0,1018.0){\rule[-0.200pt]{2.409pt}{0.400pt}}
\put(220.0,1036.0){\rule[-0.200pt]{2.409pt}{0.400pt}}
\put(1426.0,1036.0){\rule[-0.200pt]{2.409pt}{0.400pt}}
\put(220.0,1053.0){\rule[-0.200pt]{2.409pt}{0.400pt}}
\put(1426.0,1053.0){\rule[-0.200pt]{2.409pt}{0.400pt}}
\put(220.0,1067.0){\rule[-0.200pt]{4.818pt}{0.400pt}}
\put(198,1067){\makebox(0,0)[r]{1000}}
\put(1416.0,1067.0){\rule[-0.200pt]{4.818pt}{0.400pt}}
\put(220.0,113.0){\rule[-0.200pt]{0.400pt}{4.818pt}}
\put(220,68){\makebox(0,0){0}}
\put(220.0,1112.0){\rule[-0.200pt]{0.400pt}{4.818pt}}
\put(463.0,113.0){\rule[-0.200pt]{0.400pt}{4.818pt}}
\put(463,68){\makebox(0,0){0.2}}
\put(463.0,1112.0){\rule[-0.200pt]{0.400pt}{4.818pt}}
\put(706.0,113.0){\rule[-0.200pt]{0.400pt}{4.818pt}}
\put(706,68){\makebox(0,0){0.4}}
\put(706.0,1112.0){\rule[-0.200pt]{0.400pt}{4.818pt}}
\put(950.0,113.0){\rule[-0.200pt]{0.400pt}{4.818pt}}
\put(950,68){\makebox(0,0){0.6}}
\put(950.0,1112.0){\rule[-0.200pt]{0.400pt}{4.818pt}}
\put(1193.0,113.0){\rule[-0.200pt]{0.400pt}{4.818pt}}
\put(1193,68){\makebox(0,0){0.8}}
\put(1193.0,1112.0){\rule[-0.200pt]{0.400pt}{4.818pt}}
\put(1436.0,113.0){\rule[-0.200pt]{0.400pt}{4.818pt}}
\put(1436,68){\makebox(0,0){1}}
\put(1436.0,1112.0){\rule[-0.200pt]{0.400pt}{4.818pt}}
\put(220.0,113.0){\rule[-0.200pt]{292.934pt}{0.400pt}}
\put(1436.0,113.0){\rule[-0.200pt]{0.400pt}{245.477pt}}
\put(220.0,1132.0){\rule[-0.200pt]{292.934pt}{0.400pt}}
\put(45,622){\makebox(0,0){$\langle {\cal R}^2\rangle$}}
\put(828,23){\makebox(0,0){$c $}}
\put(828,1177){\makebox(0,0){ }}
\put(220.0,113.0){\rule[-0.200pt]{0.400pt}{245.477pt}}
\sbox{\plotpoint}{\rule[-0.500pt]{1.000pt}{1.000pt}}%
\multiput(226,1132)(1.336,-20.712){5}{\usebox{\plotpoint}}
\multiput(232,1039)(2.656,-20.585){5}{\usebox{\plotpoint}}
\multiput(244,946)(4.503,-20.261){2}{\usebox{\plotpoint}}
\multiput(256,892)(6.718,-19.638){2}{\usebox{\plotpoint}}
\multiput(269,854)(7.936,-19.178){2}{\usebox{\plotpoint}}
\put(289.17,808.67){\usebox{\plotpoint}}
\multiput(293,801)(11.083,-17.549){2}{\usebox{\plotpoint}}
\multiput(317,763)(13.552,-15.720){2}{\usebox{\plotpoint}}
\multiput(342,734)(15.427,-13.885){4}{\usebox{\plotpoint}}
\multiput(402,680)(17.617,-10.974){3}{\usebox{\plotpoint}}
\multiput(463,642)(18.625,-9.160){4}{\usebox{\plotpoint}}
\multiput(524,612)(19.093,-8.138){3}{\usebox{\plotpoint}}
\multiput(585,586)(19.658,-6.661){6}{\usebox{\plotpoint}}
\multiput(706,545)(20.035,-5.419){6}{\usebox{\plotpoint}}
\multiput(828,512)(20.230,-4.643){6}{\usebox{\plotpoint}}
\multiput(950,484)(20.326,-4.200){6}{\usebox{\plotpoint}}
\multiput(1071,459)(20.396,-3.845){6}{\usebox{\plotpoint}}
\multiput(1193,436)(20.450,-3.549){6}{\usebox{\plotpoint}}
\multiput(1314,415)(20.508,-3.194){6}{\usebox{\plotpoint}}
\put(1436,396){\usebox{\plotpoint}}
\put(232,1039){\makebox(0,0){$+$}}
\put(244,946){\makebox(0,0){$+$}}
\put(256,892){\makebox(0,0){$+$}}
\put(269,854){\makebox(0,0){$+$}}
\put(281,825){\makebox(0,0){$+$}}
\put(293,801){\makebox(0,0){$+$}}
\put(317,763){\makebox(0,0){$+$}}
\put(342,734){\makebox(0,0){$+$}}
\put(402,680){\makebox(0,0){$+$}}
\put(463,642){\makebox(0,0){$+$}}
\put(524,612){\makebox(0,0){$+$}}
\put(585,586){\makebox(0,0){$+$}}
\put(706,545){\makebox(0,0){$+$}}
\put(828,512){\makebox(0,0){$+$}}
\put(950,484){\makebox(0,0){$+$}}
\put(1071,459){\makebox(0,0){$+$}}
\put(1193,436){\makebox(0,0){$+$}}
\put(1314,415){\makebox(0,0){$+$}}
\put(1436,396){\makebox(0,0){$+$}}
\put(1436,396){\makebox(0,0){$+$}}
\put(1436,396){\makebox(0,0){$+$}}
\end{picture}
\end{center}
\caption{Expectation value of ${\cal R}^2$ in the ground state. Dotted
lines are connecting the data points.}
\end{figure}
wavefunction.

Knowing the wavefunctions, one can
calculate expectation values of various operators.
For example, figure 5
shows
the expectation value of ${\cal R}^2$. Even though ${\cal R}^2 >> 1$
 in the entire range of $c$, it does not mean that the nonrelativistic
approximation is valid. Recall that  $r$ is related to the dimensionless
${\cal R}$ and is measured in units of ${\cal L}$; so the expectation value of
$m r$  depends on the relation of ${\cal L}$
to the mass.

We now attempt to unwrap the dimensionless results for  B mesons.
There is not enough experimental
data, but moreover, the leading order results are too crude
to justify a precise fit. Instead we use qualitative arguments
and try to find a set of parameters that is reasonable.

We know from the heavy quark effective theory that in heavy mesons the
splittings should be independent of the heavy masses \cite{HQET}.
We also know that the spectra are not Coulombic.
Both requirements are satisfied if we choose $c$ small. For example,
with $c~=~0.1$, which is on the border of the
logarithmic regime, we find
that we can fit the splitting between the two lowest lying doublets
with $m=0.32$ GeV, $\alpha =0.35$ and
$\Lambda =0.98$ GeV. At this value of parameters
 $\tilde{\Sigma} \simeq -0.37$GeV, leading to $E_0/m \simeq 0.8$. This
shows (together with the expectation value of $mr \simeq 1.1$) that the ground
state is
not nonrelativistic, which is not surprising.

For systems consisting of lighter quarks, $c$ would be smaller because the
reduced mass decreases.
 As $c$ decreases, the state becomes more and more sensitive to the confining
potential at larger distances ${\cal R}$. At some distances
we  expect the potential to become stronger than logarithmic, but without
a calculation to a higher order in $g$ we cannot decide whether
those distances will manifest themselves in any spectra.
Also, as one deals with lighter systems, the question of restoration of
rotational symmetry becomes crucial.

\section{Summary and  conclusion. }
Starting with the canonical  light-front Hamiltonian with no zero modes, we
use a similarity transformation to find an effective Hamiltonian which is
band-diagonal with respect to a hadronic energy scale. We calculate the
effective Hamiltonian to order $g^2$ for $q\bar{q}$ color singlet states
with massive quarks of arbitrary masses. Then we split the Hamiltonian to
$H_0$, treated to all orders, and  $V$, included perturbatively,
choosing the spin-independent and rotationally invariant
 part of its nonrelativistic
reduction for $H_0$. In the nonrelativistic limit the bound state equation
leads to a
dimensionless Schr{\H{o}}dinger equation. Its scaling provides a powerful tool
to classify the operators  and estimate their expectation values.

We solve the leading order problem, and find that our calculation is
acceptable for B mesons, which we can fit with a set of reasonable,
self-consistent parameters.
We show that  heavy mesons are qualitatively different from
heavy quarkonia, but there are similarities with other
mesonic systems:  strange  mesons and isospin 1
light mesons.
 Our approach
enables us to relate different mesonic systems and use qualitative
features of spectra (e.g., almost constant mass squared splitting in the
lowest lying pseudoscalar and vector states, ordering of the lowest lying $0^+$
and $2^+$ states) as a check of our effective operators.
In this paper we present the leading order problem, study of spin and
angular momentum-dependent operators generated to this order of the
coupling by the similarity transformation will follow.
We hope that
results for heavy mesons can be generalized to lighter mesons, at least
qualitatively; although it is not clear due to  violation of
rotational symmetry.

We manage to postpone
the problem of lack of manifest rotational invariance in the confining
potential
to an order of bound state perturbation theory where there are additional
corrections beyond the current calculation.
 Nevertheless, the corrections are not
suppressed by powers of the coupling and this raises a serious warning that
we do not yet know  what is needed to recover rotational invariance. There are
several options - one might be that in our second order
light-front calculation the $q\bar{q}$ approximation is insufficient. Since
these corrections enter at the same order of bound state perturbation
theory as the emission and absorption processes, it is possible that
rotational invariance requires that a $q\bar{q} g$  component be included
to compensate for a  rotationally noninvariant $q\bar{q}$ component.
Another option is that this is not really an issue in the ``real world'' -
the states which would be mixed are well separated in energy. In this case
the presence of a rotationally noninvariant long range interactions
would be merely a nuisance because
they make it difficult to construct a simple scaling analysis that reveals
the qualitative features of meson spectra.
The resolution of these and other issues must await further calculations.

\section*{Acknowledgements}

It is a pleasure to acknowledge Ken Wilson for his involvement in this
calculation.
We are grateful for discussions throughout this work, as well as his
critical reading of the manuscript. We would also like to thank our
colleagues who took time out of their busy schedules
and carefully read the manuscript, namely Dick
Furnstahl, Dave Robertson and Kent Hornbostel. One of us (M.B.) would
like to thank Dave Robertson for his patience and numerous discussions,
Adam Szczepaniak for discussions and encouragement to pursue this work, and
Prof. Coester for bringing some formal issues to my attention.

\vfill
\pagebreak


\begin{thebibliography}{99}
\bibliographystyle{unsrt}
\bibitem{thelongpaper}{K. G. Wilson, T. S. Walhout, A. Harindranath,
Wei-Min Zhang,
R. J. Perry and S.D. G{\l}azek, Phys. Rev. {\bf D 49}, 6720 (1994),
hep-th/9401153.}

\bibitem{P2}{R. J. Perry, in  {\it Proceedings of Hadrons 94},
edited by V. Herscovitz and C. Vasconcellos (World Scientific, Singapore,
1995), hep-th/9407056.}

\bibitem{POLAND1}{K. G. Wilson and D. G. Robertson, in {\it Proc. Fourth Int.
Workshop on Light-Front Quantization and Non-Perturbative Dynamics}, edited by
S. D. G{\l}azek (World Scientific, Singapore, 1995), p. 15, hep-th/9411007. }

\bibitem{POLAND2}{K. G. Wilson and M. Brisudova, in {\it Proc. Fourth Int.
Workshop on Light-Front Quantization and Non-Perturbative Dynamics}, edited by
S. D. G{\l}azek (World Scientific, Singapore, 1995), p. 166, hep-th/9411008.}

\bibitem{similarity}{St. D. G{\l}azek and K. G. Wilson, Phys. Rev. {\bf D 48},
 5863 (1993);{\it ibid.} {\bf 49}, 4214 (1994).
A similar renormalization scheme was independently proposed by F. Wegner, Ann.
Physik {\bf 3}, 77 (1994).}

\bibitem{CQM}{S. Godfrey, N. Isgur, Phys. Rev. {\bf D 32}, 189 (1985). As
an example of more recent work see
J. F. Amundson, Phys. Rev. {\bf D 52}, 2926 (1995),
hep-ph/9504425, and references therein.}

\bibitem{robpoland}{R. J. Perry, in {\it Proc. Fourth Int.
Workshop on Light-Front Quantization and Non-Perturbative Dynamics}, edited
by S. D. G{\l}azek (World Scientific, Singapore, 1995), p.56,
 hep-th/9411037.}


\bibitem{pertren}{R. J. Perry and K. G. Wilson, Nuc. Phys. {\bf B 403}, 587
(1993).}

\bibitem{couplcoh}{R. J. Perry, Ann. Phys. {\bf 232}, 116 (1994),
hep-th/9402015.}

\bibitem{coester}{F. Coester, Prog. in Part. Nucl. Phys. {\bf 29}, 1, (1992).}

\bibitem{xy}{W. M. Zhang and A. Harindranath, Phys. Rev. {\bf D 48}, 4868
(1993); {\it ibid.} 4881; {\it ibid.} 4903.}

\bibitem{lebro}{G. P. Lepage, S. J. Brodsky, Phys. Rev. {\bf D 22}, 2157
(1980).}

\bibitem{bass}{A. Bassetto, G. Nardelli, R. Soldati, {\it Yang-Mills
Theories in Algebraic Non-covariant Gauges} (World Scientific, Singapore,
1991).}

\bibitem{HQET}
{J. M. Flynn, N. Isgur, J. Phys. {\bf G 18}, 1627 (1992);
N. Isgur, M. Wise, Phys. Rev. Lett. {\bf 66}, 1130 (1991);
G. P. Lepage, B. A. Thacker, in: Field theory on the lattice,
Proc.~Intern.~Symp. (Seillac, France,~1987), eds. A. Billoire et al.,
Nucl.~Phys.~{\bf B}(Proc., Suppl.) {\bf 4}, 199 (1988);
H. Georgi, Phys. Lett. {\bf B 240}, 447 (1990). For a recent review on the HQET
see M. Neubert, Phys. Rept. {\bf 245}, 259 (1994).
}
\end{thebibliography}
\end{document}